\documentclass[prd,tightenlines,nofootinbib,superscriptaddress]{revtex4}

\usepackage{pre_all}
\usepackage{pre_CS}
\allowdisplaybreaks

\hypersetup{
pdfstartview = {FitH},
}
\hypersetup{
	colorlinks=true,         
	linkcolor=brown,          
	citecolor=red,        
	urlcolor=blue            
}

\begin{document}

\title{Geometrical Reconstruction of Spinfoam Critical Points with A Cosmological Constant}
\author{{\bf Qiaoyin Pan}}\email{qpan@fau.edu}
\affiliation{Department of Physics, Florida Atlantic University, 777 Glades Road, Boca Raton, FL 33431, USA}

\date{\today}

\begin{abstract}
 
In this work, we present a geometrical reconstruction of the critical points of the spinfoam amplitude for a 4D Lorentzian model with a non-zero cosmological constant. By establishing the correspondence between the moduli space of $\SL(2,\bC)$ flat connections on the graph-complement 3-manifold $\SG$ and the geometry of a constantly curved 4-simplex, we demonstrate how the critical points encode discrete curved geometries. The analysis extends to 4-complexes dual to colored graphs, aligning with the improved spinfoam model recently introduced in \cite{Han:2025abc2}. Central to this reconstruction are translating the geometry of constantly curved 4-simplices into Fock-Goncharov coordinates and spinors, which translate the geometry data into holonomies and symplectic structures, thereby defining the critical points of the spinfoam amplitude. This framework provides an algorithmic foundation for computing quantum gravity corrections and opens avenues for applications in quantum cosmology and black hole physics, where the cosmological constant plays a pivotal role.

\end{abstract}

\maketitle

\tableofcontents

\section{Introduction}

In the study of covariant loop quantum gravity (LQG), also called the spinfoam model \cite{Rovelli:2014ssa,Perez:2012wv}, the semi-classical regime ($\hbar\rightarrow 0$) is one of the focuses of attention as it is where consistency with general relativity is tested and quantum corrections are extracted. A well-defined spinfoam model on the triangulated spacetime should reproduce, at the semi-classical regime, the Regge action of this triangulation, which is the discretized Einstein gravity. This has been known to be true for the Engle-Pereira-Rovelli-Livine (EPRL) model \cite{Engle:2007wy} and Freidel-Krasnov (FK) model \cite{Freidel:2007py}, which both describe transition amplitudes on four-dimensional spacetime with a vanishing cosmological constant $\Lambda$. 
Technically speaking, the semi-classical approximation relies on the stationary point analysis of the spinfoam amplitude, and the Regge action appears at the critical points \cite{Barrett:1998gs,Barrett:2009mw,Conrady:2008mk}. We, therefore, say that the critical points of spinfoam model with $\Lambda=0$ correspond to (discrete) flat geometry. 

The critical point properties have been recently shown to be generalized to the Lorentzian spinfoam model with a non-zero $\Lambda$ firstly introduced in \cite{Han:2021tzw} and further studied in \cite{Han:2023hbe,Han:2024reo,Han:2025abc2}. 
Namely, the stationary phase analysis can be applied to the semi-classical approximation of the spinfoam amplitude as an oscillatory action shows up at the exponent of the integrand in this approximation. The critical points reproduce the Regge action for constantly curved 4-complex, which means they correspond to (discrete) constantly curved geometry \cite{Han:2024reo,Han:2025abc2}. 
Other welcoming properties of this spinfoam model include that the spinfoam amplitude for any spacetime triangulation is finite by construction, and that the amplitude can be explicitly realized using Chern-Simons phase space coordinates on a 3-manifold generated by the spacetime 4-manifold and a particular graph. 

This spinfoam model is the center of study of this paper. 
We are interested in the realization of the semi-classical approximation of the spinfoam amplitude for a general 4-complex. Given a 4-complex that is composed of a large number of 4-simplices, the expression of the amplitude takes an involved form as it is built with numerous vertex amplitudes, edge amplitudes and face amplitudes, each of which takes a non-trivial form. 
It is, practically speaking, a non-trivial task to extract the critical points from the stationary phase analysis for the spinfoam amplitude. 
However, as we have learnt from formal analysis \cite{Han:2025abc2} that each critical point gives rise to a constantly curved geometry with fixed boundary conditions, one can go over the stationary phase analysis can reproduce the critical points directly from the geometry data of a constantly curved 4-complex, which in turn computes the semi-classical approximation of the amplitude at the zero-order. 
We call this process the geometrical reconstruction of the spinfoam critical points. 
This is also the spirit behind the series of work \cite{Han:2020fil,Han:2023cen,Han:2024lti,Li:2025nog} on the numerical realization of the leading order and next-to-leading order spinfoam amplitudes for EPRL model at large-$j$ approximation.  

In this paper, we study the spinfoam model recently defined in \cite{Han:2025abc2}, which is an improved version of the one introduced in \cite{Han:2021tzw} and studied in \cite{Han:2023hbe,Han:2024reo}. In the improved model, the vertex amplitude is defined differently compared to \cite{Han:2021tzw} by using a new set of phase space coordinates, making it possible to define a simple and universal face amplitude. 
The new vertex amplitude is defined based on the Chern-Simons theory with level $k\in 8\Z_+$, which is related to the value of the cosmological constant $\Lambda$, on a graph-complement of 3-sphere $S^3$. The spinfoam amplitude is generalized to spinfoam graphs corresponding to colored graphs used in the colored tensor model \cite{Gurau:2011aq,Gurau:2011xp}. 
 For these spinfoam amplitudes, it has also been shown in \cite{Han:2025abc2} that the semi-classical approximation ($k\rightarrow\infty$) of the improved model can be computed systematically. 

This paper is organized as follows. We first discuss why 4D constantly curved geometry can be implemented in moduli space of flat connection on a 3-manifold, where Chern-Simons theory is defined. This sets the basis for using Chern-Simons theory on this 3-manifold to construct spinfoam amplitude on a 4-simplex. 
In particular, in Sections \ref{sec:flat_connection_4_simplex}, we review the holonomy representation of the fundamental group of the 1-skeleton of a 4-simplex and the isomorphic fundamental group of a related 3-manifold. Using this isomorphism, Section \ref{sec:flat_connection_2_geometry} delves into extracting the 4-simplex geometries from the holonomies defined by the Chern-Simons theory. 
In Sections \ref{sec:review_SF}, we review the construction of the spinfoam amplitudes and their critical points introduced in \cite{Han:2025abc2}. 
In Section \ref{sec:geo_reconstruct}, we reproduce these critical points using the geometrical data extracted in the Section \ref{sec:flat_connection_4_simplex} --\ref{sec:flat_connection_2_geometry}. 
We conclude and discuss possible applications of this framework in Section \ref{sec:conclusion}.

\section{Flat connection on 3-manifold and curved 4-simplex geometry}
\label{sec:flat_connection_4_simplex}

A 4-simplex, topologically isomorphic to a 4-ball, is bounded by 5 tetrahedra whose boundary triangles are glued together pairwise. In this paper, we consider oriented convex constantly curved 4-simplex bounded by constantly curved tetrahedra whose global curvature can be positive or negative and the curvature $R$ is given by the value of the cosmological constant: $R=\nu\sqrt{3/|\Lambda|}$ with $\nu:=\sgn(\Lambda)$. 

The geometry of a constantly curved convex tetrahedron can be uniquely determined (up to orientation) by four $\SU(2)$ group elements, denoted by $\{H_i\}_{i=1,\cdots,4}$, which satisfy the {\it closure condition}:
\be
H_4H_3H_2H_1=\id_{\SU(2)}\,.
\label{eq:closure_tetra}
\ee
\begin{figure}[h!]
\begin{subfigure}[t]{0.2\linewidth}
\includegraphics{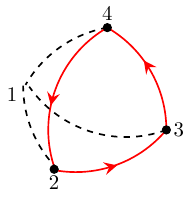}
\caption{$p_1$}
\label{fig:ciliated_graph}
\end{subfigure}
\begin{subfigure}[t]{0.2\linewidth}
\includegraphics{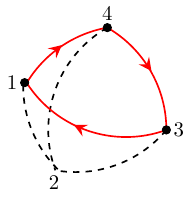}
 \caption{$p_2$}
\label{fig:ciliated_graph}
\end{subfigure}
\begin{subfigure}[t]{0.2\linewidth}
\includegraphics{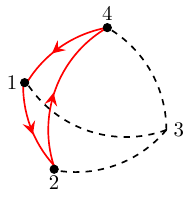}
 \caption{$p_3$}
\label{fig:ciliated_graph}
\end{subfigure}
\begin{subfigure}[t]{0.2\linewidth}
\includegraphics{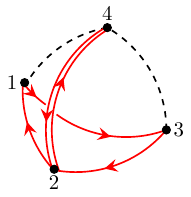}
 \caption{$p_4$}
\label{fig:simple_path}
\end{subfigure}
\caption{The set of simple paths ({\it in red}) for holonomies $\{H_1,H_2,H_3,H_4\}$ defined with vertex 4 as the base point and the edge connecting vertices 2 and 4 as the special edge. They satisfy the closure condition \eqref{eq:closure_tetra}. Simple paths on a tetrahedron with negative curvature are defined similarly. }
\label{fig:simple_paths}
\end{figure}
This is called the {\it curved Minkowski Theorem} proven in \cite{Haggard:2015ima}. 
Each $H_i=H_i(\omega_{\rm spin})$ describes the holonomy of the spin connection $\omega_{\rm spin}$ along a simple path $p_i$ as shown in fig.\ref{fig:simple_paths} based at the same vertex. 
The fundamental group of the one-skeleton of a tetrahedron is constructed by these simple paths
\be
\pi_1({\rm sk}_1(\tetra))=\{p_1,p_2,p_3,p_4|p_4\circ p_3\circ p_2\circ p_1=1\}\,,
\ee
and it is isomorphic to the fundamental group $\pi_1(\Sfour)$ of a 4-holed sphere $\Sfour$. This isomorphism, denoted as $X:\pi_1({\rm sk}_1(\tetra))\rightarrow \pi_1(\Sfour)$, leads to the fact that a flat connection in the moduli space $\cM_\Flat(\Sfour,\SU(2))\equiv \Hom(\pi_1(\Sfour),\SU(2))/\SU(2)$ of $\SU(2)$ flat connection on $\Sfour$ one-to-one corresponds to the geometry of a tetrahedron described by the holonomies of $\SU(2)$ spin connection along the simple paths of the tetrahedron, as shown in the following diagram.
\be
\ba{c}
\pi_1({\rm sk}_1(\tetra))\quad \xrightarrow{\phantom{yyyy}X\phantom{yyyy}} \quad \pi_1(\Sfour)	\\\\
\omega_{\rm spin}\searrow\qquad\qquad\qquad\qquad \swarrow \omega_{\Flat}\\\\
\{H_1,H_2,H_3,H_4\in\SU(2)|H_4H_3H_2H_1=\id_{\SU(2)}\}/\SU(2)\,,
\ea
\label{eq:iso_tetra_sphere}
\ee
where the quotient is by the conjugate action of $\SU(2)$. 

Such an isomorphism can be generalized to a one-higher dimensional case \cite{Haggard:2015nat}. 
To rephrase, \eqref{eq:iso_tetra_sphere} relates the fundamental group of a 3-simplex, \ie a tetrahedron, and that of the nodes-complement of its topological boundary $S^2$ where the nodes are the (3-3=) 0-subcomplexes of the dual 2-complex on the boundary of the 3-simplex. Its generalization gives the isomorphism between the fundamental group of (the one-skeleton of) a 4-simplex and that of the graph-complement of its topological boundary $S^3$ where the graph is the (4-3=)1-subcomplex -- $\Gamma_5$ graph -- of the dual 3-complex on the boundary of the 4-simplex. Denote this graph-complement 3-manifold as $\SG$.

\begin{figure}[h!]
\centering
\includegraphics[width=0.25\textwidth]{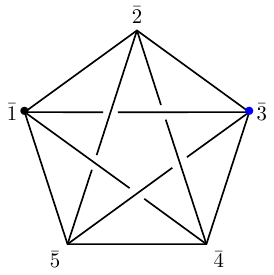}
\caption{A 4-simplex projected on $\R^2$. Numbers $\bar{1},\cdots,\bar{5}$ denote the vertices. $\tetra_a$ denotes the tetrahedron that does not contain the vertex $\bar{a}$. $\triangle_{ab}\equiv\triangle_{ba}$ denotes the triangle shared by $\tetra_a$ and $\tetra_b$. The over- and under-crossing specify the correct relative positions of vertices in each tetrahedron.} 
\label{fig:4-simplex}
\end{figure}

To write this isomorphism exactly, let us specify the fundamental group of a 4-simplex and $\SG$ separately. The generators of the former are the closed paths based at the same vertex along the 1-skeleton and circling around a triangle. We refer to fig.\ref{fig:4-simplex} and fix the notations as follows. 
We use numbers $\bar{1},\cdots,\bar{5}$ with bars to denote the vertices of the 4-simplex and $(\bar{a}\bar{b})$ to denote the oriented edge that connects source $\bar{b}$ to target $\bar{a}$. Denote $(\bar{b}\bar{a})=(\bar{a}\bar{b})^{-1}$. $\tetra_a$ denotes the tetrahedron that does not contain the vertex $\bar{a}$. Each pair of tetrahedra $\tetra_a$ and $\tetra_b$ share a triangle $\triangle_{ab}$ (or $\triangle_{ba}$), which is the one that does not contain vertices $\bar{a}$ and $\bar{b}$. 

We choose $\bar{1}$ to be the base point. $p_{ab}$ denotes the oriented closed path based at $\bar{1}$ that circles $\triangle_{ab}$ and whose orientation matches the outgoing normal of $\triangle_{ab}$ in $\tetra_a$. To fix the path for triangles not attached to $\bar{1}$, which is the case for all triangles in $\tetra_1$, we need to additionally specify a ``special edge'' that connects $\bar{1}$ to a vertex on the boundary of the triangle. Two special edges are needed at the minimum. We choose $(\bar{3}\bar{1})$ to be the special edge for triangles $\triangle_{12},\triangle_{14},\triangle_{15}$ and choose $(\bar{5}\bar{1})$ to be the special edge for triangle $\triangle_{31}$ on $\tetra_3$. For instance, $p_{12}=(\bar{1}\bar{3})\circ (\bar{3}\bar{5})\circ (\bar{5}\bar{4})\circ (\bar{4}\bar{3})\circ (\bar{3}\bar{1})$. $p_{ba}=p_{ab}^{-1}$ holds for all $(\bar{a}\bar{b})\neq (\bar{1}\bar{3})$ or $(\bar{3}\bar{1})$. Specially, 
\be\ba{l}
p_{13}:=(\bar{1}\bar{3})\circ(\bar{3}\bar{5})\circ (\bar{5}\bar{2})\circ (\bar{2}\bar{4})\circ (\bar{4}\bar{5})\circ (\bar{5}\bar{3})\circ(\bar{3}\bar{1})\,,\\[0.15cm]
p_{31}:=(\bar{1}\bar{5})\circ (\bar{5}\bar{4})\circ (\bar{4}\bar{2})\circ (\bar{2}\bar{5})\circ (\bar{5}\bar{1})\,.
\ea\ee 
Therefore, $p_{13}$ and $p_{31}$ are related by
\be
p_{13}=p_{24}\circ p_{31}^{-1}\circ p_{24}^{-1}\,.
\label{eq:p13_to_p31}
\ee

The generators of the fundamental group $\pi_1({\rm sk}_1((\text{4-simplex}))$ of a 4-simplex are then given by the following 5 relations.
\begin{subequations}
\begin{align}
\tetra_1:\quad & p_{13}\circ p_{12}\circ p_{15}\circ p_{14}=1\,,\\
\tetra_2:\quad & p_{12}^{-1}\circ p_{24}\circ p_{23}\circ p_{25}=1\,,\\
\tetra_3:\quad & p_{31}\circ p_{34}\circ p_{35}\circ p_{23}^{-1}=1\,,\\
\tetra_4:\quad & p_{14}^{-1}\circ p_{45}\circ p_{34}^{-1}\circ p_{24}^{-1}=1\,,\\
\tetra_5:\quad & p_{15}^{-1}\circ p_{12}^{-1}\circ p_{35}^{-1}\circ p_{45}^{-1}=1\,.
\end{align}
\label{eq:fundamental_group_4simplex}
\end{subequations}
That is, $\pi_1({\rm sk}_1(\text{4-simplex}))=\{\{p_{ab}\}_{a\neq b}|{\rm Eqns.}\eqref{eq:p13_to_p31} - \eqref{eq:fundamental_group_4simplex}\}$.

\begin{figure}[h!]
\centering
\includegraphics[width=0.3\textwidth]{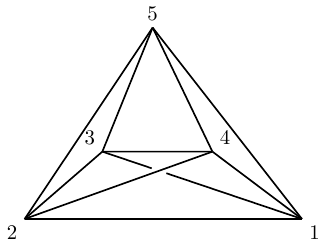}
\caption{$\Gamma_5$ graph projected on $\R^2$. }
\label{fig:Gamma5}
\end{figure}

On the other hand, the fundamental group of $\SG$ can be computed by a generalized Wirtinger representation \cite{brown2006topology}. It is done in the following steps. Firstly, project $\Gamma_5$ onto a plane as in fig.\ref{fig:Gamma5}. Denote the nodes of $\Gamma_5$ by numbers $1,\cdots,5$ (with no bars) and the oriented link connecting the target node $a$ and source node $b$ by $e_{ab}$. There is one crossing that breaks link $e_{13}$ into two links, denoted as $e_{13}^{(1)}$ for the one attached to vertex $1$ and $e_{13}^{(3)}$ for the one attached to vertex $3$, so there are totally 11 links under this projection, each is associated with a fundamental group generator of $\SG$. Choose a base point $\fb$ in $\SG$. The generator associated to $e_{ab}$ is given by a non-contractible closed loop $\fl_{ab}$ based at $\fb$ circling $e_{ab}$ whose orientation matches that of $e_{ab}$. Specifically, the generators associated to $e_{13}^{(1)}$ and $e_{13}^{(3)}$ respectively are denoted as $\fl_{13}^{(1)}$ and $\fl_{13}^{(3)}$ respectively. 
We associate an orientation to each $\fl_{ab}$ such that it matches the orientation of $e_{ab}$. Then $\fl_{ba}=\fl_{ab}^{-1}$ for $(a,b)\neq (1,3)$ or $(3,1)$. The 11 generators are subject to the following relations, one for each node or crossing.
\begin{subequations}
\begin{align}
\text{node 1}:\quad & \fl_{13}^{(1)}\circ \fl_{12}\circ \fl_{15}\circ \fl_{14}=1\,,\\
\text{node 2}:\quad & \fl_{12}^{-1}\circ \fl_{24}\circ \fl_{23}\circ \fl_{25}=1\,,\\
\text{node 3}:\quad & \fl_{13}^{(3)\,-1}\circ \fl_{34}\circ \fl_{35}\circ\fl_{23}^{-1}=1\,,\\
\text{node 4}:\quad & \fl_{14}^{-1}\circ \fl_{45}\circ \fl_{34}^{-1}\circ \fl_{24}^{-1}=1\,,\\
\text{node 5}:\quad & \fl_{15}^{-1}\circ \fl_{12}^{-1}\circ \fl_{35}^{-1}\circ \fl_{45}^{-1}=1\,,\\
\text{crossing}:\quad & \fl_{13}^{(1)}=\fl_{24}\circ \fl_{13}^{(3)}\circ \fl_{24}^{-1}\,.
\end{align}
\label{eq:fundamental_group_3manifold}
\end{subequations}
Therefore, $\pi_1(\SG)=\{\{\fl_{ab}\}_{a\neq b}|{\rm Eqn.}\eqref{eq:fundamental_group_3manifold}\}$. 
Already from the definition, one can immediately notice an isomorphism $Y:\pi_1({\rm sk}_1(\text{4-simplex}))\rightarrow\pi_1(\SG)$ that maps $Y(p_{ab})=\fl_{ab}$ for $(\bar{a}\bar{b})\neq (\bar{1}\bar{3})$ or $(\bar{3}\bar{1})$ and $Y(p_{13})=\fl_{13}^{(1)},\,Y(p_{31})=\fl_{13}^{(3)\,-1}$. Given a representation $\rho=\Hom(\pi_1(\SG),\SL(2,\bC))$ such that $\rho(\fl_{ab})=\Ht_{ab}$ and that $\rho(\fl_{ab}^{-1})=\Ht_{ab}^{-1}$, \eqref{eq:fundamental_group_3manifold} gives 5 closure conditions on the holonomies and a conjugate relation to $\Ht^{(1)}_{13}$ and $\Ht^{(3)}_{13}$: 
\begin{subequations}
\begin{align}
& \Ht_{13}^{(1)}\Ht_{12}\Ht_{15}\Ht_{14}=\id\,,
\label{eq:closure_3manifold_1}\\
& \Ht_{12}^{-1}\Ht_{24}\Ht_{23}\Ht_{25}=\id\,,\\
& \Ht_{13}^{(3)\,-1}\Ht_{34}\Ht_{35}\Ht_{23}^{-1}=\id\,,\\
& \Ht_{14}^{-1}\Ht_{45}\Ht_{34}^{-1}\Ht_{24}^{-1}=\id\,,\\
& \Ht_{15}^{-1}\Ht_{25}^{-1}\Ht_{35}^{-1}\Ht_{45}^{-1}=\id\,,
\label{eq:closure_3manifold_5}\\
& \Ht_{13}^{(1)}=\Ht_{24}\Ht_{13}^{(3)}\Ht_{24}^{-1}\,.
\label{eq:closure_3manifold_6}
\end{align}
\label{eq:closure_3manifold}
\end{subequations}
$\Ht_{ba}=\Ht_{ab}^{-1}$ for $(a,b)\neq(1,3)$ or $(3,1)$. 
Representing $\pi_1({\rm sk}_1(\text{4-simplex}))$ also in $\SL(2,\bC)$ by $\rho'=\Hom(\pi_1({\rm sk}_1(\text{4-simplex})),\SL(2,\bC))$ and identifying $\rho'(p_{ab})=\rho(\fl_{ab})$ for all $(a,b)\neq (1,3)$ or $(1,3)$ while $\rho'(p_{13})=\rho(\fl_{13}^{(1)})$ and $\rho'(p_{31})=\rho(\fl_{13}^{(3)\,-1})$, \eqref{eq:closure_3manifold_1}--\eqref{eq:closure_3manifold_5} are nothing but the 5 copies of closure conditions as in \eqref{eq:closure_tetra} represented in $\SL(2,\bC)$, each corresponding to a tetrahedron on the boundary of the 4-simplex and \eqref{eq:closure_3manifold_6} relates the holonomy $H_{13}^{(1)}$ around $\triangle_{13}$ as the boundary of $\tetra_1$ and the holonomy $H_{13}^{(3)}$ around the same triangle as the boundary of $\tetra_3$. $\rho$ and $\rho'$ effectively associate connections $\omega_\Flat$ and $\omega_{\text{spin}}$ respectively to the 4-simplex and $\SG$ respectively. We then have a similar commuting map as \eqref{eq:iso_tetra_sphere} but in one higher dimension represented in $\SL(2,\bC)$. 
\be
\ba{c}
\pi_1({\rm sk}_1(\text{4-simplex}))\quad \xrightarrow{\phantom{yyyy}Y\phantom{yyyy}} \quad \pi_1(\SG)	\\\\
\omega_{\rm spin}\searrow\qquad\qquad\qquad\qquad \swarrow \omega_{\Flat}\\\\
\{\{\Ht_{ab}\}\in\SL(2,\bC)|\text{Eqn.}\eqref{eq:closure_3manifold}\}/\SL(2,\bC)\,,
\ea
\label{eq:iso_simplex_manifold}
\ee
where the quotient is by the conjugate action of $\SL(2,\bC)$. 

Note that \eqref{eq:closure_3manifold} does not define the geometry of a 4-simplex as $\Ht_{ab}\in\SL(2,\bC)$ instead of $\SU(2)$. However, $\Ht_{ab}\in\SU(2)\,,\forall\,(\bar{a}\bar{b})$ would be a too strong requirement for identifying the geometry of a 4-simplex. We will see in the next section that a looser restriction, which will be shown to be given by the spinfoam boundary condition in Section \ref{sec:geo_reconstruct}, is enough for reconstructing the geometry of a 4-simplex from \eqref{eq:closure_3manifold}. 

\section{From flat connection to curved geometry}
\label{sec:flat_connection_2_geometry}

Let us now specify how imposing restrictions on the holonomies $\{\Ht_{ab}\}$ allows us to describe the geometry of a constantly curved 4-simplex.
We require that each $\SL(2,\bC)$ holonomy $\Ht_{ab}$ of the flat connection defined in \eqref{eq:closure_3manifold} is restricted to that conjugate to an $\SU(2)$ holonomy, denoted as $O_{ab}$, and its eigenvalue gives the area $\fa_{ab}$ of $\triangle_{ab}$. More precisely, let 
\be\ba{l}
\Ht_{ab}=g_aO_{ab}g_a^{-1} = g_bO^{-1}_{ba}g_b^{-1}\,,\quad (a,b)\neq (1,3),(3,1)\,,\\[0.15cm]
\Ht_{13}^{(1)}=g_1O_{13}g_1^{-1}\,,\quad
\Ht_{13}^{(3)\,-1}=g_3O_{31}g_3^{-1}\,.
\ea
\label{eq:Ht_to_O}
\ee
where $g_a,g_b\in\SL(2,\bC)$ and $O_{ab},O_{ba}\in\SU(2)$. 
$g_a$ can be geometrically interpreted as parallel transporting the base point $\fb$ in $\SG$ to the base point $\fb_a$ on the 4-holed sphere $\cS_a\subset \partial(\SG)$ defining $\{O_{ac}\}_{c}$. Then $O_{ab}$ is the holonomies on $\cS_a$ based at $\fb_a$ around the hole as a boundary of the annulus connecting to $\cS_b$, denoted as $(ab)$\footnote{We remind the readers that $(ab)$ should not be confused with $(\bar{a}\bar{b})$. The former denotes the 2-dimensional annulus connecting 4-holed spheres $\cS_a$ and $\cS_b$ while the latter denotes the edge in the 4-simplex that connects vertices $\bar{a}$ and $\bar{b}$. }, oriented in the outgoing direction of $\cS_a$. 
In this way, the constrained versions of closure conditions \eqref{eq:closure_3manifold_1}--\eqref{eq:closure_3manifold_5} are
\begin{subequations}
\begin{align}
& O_{13}O_{12}O_{15}O_{14}=\id\,,\\
& O_{21}O_{24}O_{23}O_{25}=\id\,,\\
& O_{31}O_{34}O_{35}O_{32}=\id\,,\\
& O_{41}O_{45}O_{43}O_{42}=\id\,,\\
& O_{51}O_{52}O_{53}O_{54}=\id\,.
\end{align}
\label{eq:closure_5spheres}
\end{subequations}
These $\SU(2)$ holonomies are subject to the constraints
\be
O_{ab}=G_{ab}O_{ba}^{-1}G_{ba}\,,\quad 
G_{ba}=G_{ab}^{-1}\in\SL(2,\bC)\,,\quad \forall\,(a,b)
\label{eq:constraints_on_Oab}
\ee
where
\be\ba{l}
G_{ab}:=g_a^{-1}g_b\,,\quad \forall \, (a,b)\neq (1,3),(3,1)\,,\\[0.15cm]
G_{13}:=g_1^{-1}\lb g_2O_{24}g_2^{-1} \rb g_3=G_{31}^{-1}\,.
\ea
\label{eq:def_Gab}
\ee
$G_{ab}$ then represents the parallel transport from $\fb_b$ to $\fb_a$ along a path passing through the common base point $\fb$ in $\SG$. In other words, it changes the local frame from $\tetra_b$ to $\tetra_a$ and thus we call it a {\it frame-changing holonomy}. 
The second line of \eqref{eq:def_Gab} together with \eqref{eq:constraints_on_Oab} is the constrained version of \eqref{eq:closure_3manifold_6}.

As an $\SU(2)$ element, $O_{ab}$ can be factorized as follows.
\be
O_{ab}=M(\xi_{ab})\diag(\lambda_{ab},\lambda_{ab}^{-1}) M(\xi_{ab})^{-1}\,,\,\,
\lambda_{ab}=
e^{-i\f{|\Lambda|}{6}\fa_{ab}}\in{\rm U}(1)\,,
\label{eq:Oab_decomp}
\ee
where $M(\xi_{ab})$ is defined in terms of a spinor $|\xi_{ab}\ra=(\xi_{ab}^0,\xi_{ab}^1)^\top\in\bC^2$ and its dual spinor $|\xi_{ab}]=(-\bar{\xi}_{ab}^1,\bar{\xi}_{ab}^0)^\top$ assigned to the hole of $\cS_a$ that connects to $\cS_b$. 
$|\xi_{ab}\ra$ and $|\xi_{ab}]$ are indeed eigenvectors of $O_{ab}$. 
$|\xi_{ab}]$ is dual to $|\xi_{ab}\ra$ in the sense that $[\xi_{ab}|\xi_{ab}\ra=\la \xi_{ab}|\xi_{ab}]=0$ (by definition). They further satisfy the normalization property $\la \xi_{ab}|\xi_{ab}\ra:=\bar{\xi}_{ab}^0\xi_{ab}^0+\bar{\xi}_{ab}^1\xi_{ab}^1=1=[\xi_{ab}|\xi_{ab}]$ which guarantees that $M(\xi_{ab})\in\SU(2)$ by the following definition.
\be
M(\xi_{ab}):=\mat{cc}{|\xi_{ab}\ra \,,& |\xi_{ab}]}=\mat{cc}{\xi_{ab}^0 & -\bar{\xi}_{ab}^{1}\\[0.1cm] \xi_{ab}^1 & \bar{\xi}_{ab}^0}
\ee
As a headsup, since $|\xi_{ab}\ra$ is the normalized eigenvector of $O_{ab}$ at $\fb_a$, we will see in Section \ref{sec:geo_reconstruct} that it can be coordinates on $\cM_{\Flat}(\cS_a,\SU(2))$. 

Recalling the isomorphism \eqref{eq:iso_tetra_sphere} between the moduli space of flat connection on a 4-holed sphere and the geometry of a tetrahedron, the geometry of $\tetra_a$ is encoded in $\{O_{ab}\}_{b}$: in the decomposition \eqref{eq:Oab_decomp}, $\lambda_{ab}$ encodes the area $\fa_{ab}=\fa_{ba}$ of $\triangle_{ab}$ and $|\xi_{ab}\ra$ encodes the 3D normal vector to $\triangle_{ab}$ in the local frame of $\tetra_a$.
By convexity, $\fa_{ab}$ is bounded: $\fa_{ab}\in [0,6\pi/|\Lambda|]$ \footnote{We consider the areas calculated with the Gau\ss-Bonnet theorem, so the areas of hyperbolic triangles are also bounded. See more details in \cite{Haggard:2015ima}. }. The normal $\hat{n}_{ab}$ is then
\be
\hat{n}_{ab}=
\begin{cases}
\la\xi_{ab}| \vec{\sigma}| \xi_{ab}\ra\,,\quad & \text{if } \fa_{ab}\in [0,3\pi/|\Lambda|)\\
-\la\xi_{ab}| \vec{\sigma}| \xi_{ab}\ra\,,\quad& \text{if } \fa_{ab}\in [3\pi/|\Lambda|,6\pi/|\Lambda|)
\end{cases}\,,
\label{eq:normal_ab}
\ee
where $\vec{\sigma}=(\sigma^1,\sigma^2,\sigma^3)$ is a vector of Pauli matrices. Inversely, one can solve for $\lb\xi^0_{ab},\xi^1_{ab}\rb$ upon a gauge-fixing (since $\lb\xi^0_{ab},\xi^1_{ab}\rb\rightarrow \lb e^{i\theta}\xi^0_{ab},e^{i\theta}\xi^1_{ab}\rb$ is a ${\rm U}(1)$ gauge transformation for any $\theta\in\R$).
\be
\xi_{ab}^0=\begin{cases}
\sqrt{\f{1+n_{ab}^3}{2}}\,,\quad & \text{if } \fa_{ab}\in [0,3\pi/|\Lambda|)\\
\sqrt{\f{1-n_{ab}^3}{2}}\,,\quad & \text{if } \fa_{ab}\in [3\pi/|\Lambda|,6\pi/|\Lambda|)
\end{cases}\,,\,\,
\xi_{ab}^1=\begin{cases}
	\sqrt{\f{1-(n_{ab}^3)^2}{2(1+n_{ab}^3)}}e^{i\psi}\,,\quad & \text{if }\fa_{ab}\in [0,3\pi/|\Lambda|)\\
	\sqrt{\f{1-(n_{ab}^3)^2}{2(1-n_{ab}^3)}}e^{i\psi}\,,\quad & \text{if }\fa_{ab}\in [3\pi/|\Lambda|,6\pi/|\Lambda|)
\end{cases}\,,
\,\,
\psi=\arctan\lb\f{ n^2_{ab}}{n^1_{ab}}\rb\,.
\label{eq:xi_from_n}
\ee
The outward-pointing normal $\hfn_{ab}$ to $t_i^{(a)}$ is different from $\hat{n}_{ab}$ by the sign $\nu$ of $\Lambda$:
\be
\hfn_{ab}=\nu \hat{n}_{ab}\,.
\label{eq:outward_normal}
\ee
On the other hand, a similar factorization for $O_{ba}$ gives
\be
O_{ba}=M(\xi_{ba})\diag(\lambda_{ba},\lambda_{ba}^{-1}) M(\xi_{ba})^{-1}\,,\,\,
\lambda_{ba}=e^{i\f{|\Lambda|}{6}\fa_{ab}}\,,
\label{eq:Oba_decomp}
\ee
where  $\lambda_{ba}=\lambda_{ab}^{-1}$ and $M(\xi_{ba})$ is defined in the same way as $M(\xi_{ab})$ but with spinors $|\xi_{ba}\ra$ and its dual $|\xi_{ba}]$ as eigenvectors of $O_{ba}$ at $\fb_b$ on $\cS_b$. Importantly, the 3D normal vector to $\triangle_{ab}$ in the local frame of $\tetra_b$ defined as 
\be
\hat{n}_{ba}=
\begin{cases}
\la\xi_{ba}| \vec{\sigma}| \xi_{ba}\ra\,,\quad & \text{if }\fa_{ab}\in [0,3\pi/|\Lambda|)\\
-\la\xi_{ba}| \vec{\sigma}| \xi_{ba}\ra& \text{if }\fa_{ab}\in [3\pi/|\Lambda|,6\pi/|\Lambda|)
\end{cases}
\label{eq:normal_ba}
\ee
is different from $\hat{n}_{ab}$ in general as the two spinors are different. 
Indeed, $\hat{n}_{ab}$ and $\hat{n}_{ab}$ are related by the dihedral angle, denoted as $\Theta_{ab}$ of $\tetra_a$ and $\tetra_b$ hinged by $\triangle_{ab}$. 
$\Theta_{ab}$ is encoded in the frame-changing holonomy $G_{ab}$ and the pair of spinors $\lb |\xi_{ab}\ra,|\xi_{ba}\ra\rb$ (or $\lb |\xi_{ab}],|\xi_{ba}]\rb$):
\be
G_{ab}=M(\xi_{ab})\mat{cc}{ \gamma_{ab}&0\\0&\gamma_{ab}^{-1}}M(\xi_{ba})^{-1}\,,\,\,
\gamma_{ab}=e^{-\nu\sgn(V_4)\f{\Theta_{ab}}{2}+i\theta_{ab}}\,,
\label{eq:G_factorize}
\ee
where $\Theta_{ab},\theta_{ab}\in\R$ and $V_4$ is the volume of the 4-simplex. $\theta_{ab}$ is a non-geometrical parameter. Then $|\xi_{ab}\ra$ (\resp $|\xi_{ab}]$) and $|\xi_{ba}\ra$ (\resp $|\xi_{ba}]$) are related by a simple parallel transport:
\be
|\xi_{a b}\rangle=\gamma_{a b}^{-1} G_{a b}|\xi_{b a}\rangle\,, \quad |\xi_{a b}]=\gamma_{a b} G_{a b}|\xi_{b a}]\,.
\label{eq:xi_G_relation}
\ee
Therefore, given the geometry of a 4-simplex, including the areas and normals of all triangles in different tetrahedron frames, one can reconstruct all the $G_{ab}$'s up to some phases $\{\theta_{ab}\}_{a\neq b}$ determined by the boundary condition (as all edges of a 4-simplex are on the boundary). 
Further, flat connection holonomies $\{\Ht_{ab}\}$ on $\SG$ can be determined by $\{G_{ab}\}$ through \eqref{eq:Ht_to_O} up to an $\SL(2,\bC)$ gauge as $G_{ab}$ is invariant under the gauge transformation from the left $g_a\rightarrow hg_a\,,\,\forall h\in\SL(2,\bC)$ ({\it r.f. }\eqref{eq:def_Gab}). Such a gauge transformation corresponds to changing the common base point $\fb\rightarrow\fb'$ for defining $\{\Ht_{ab}\}$.

\section{Review of the spinfoam model with $\Lambda\neq 0$ and the critical points}
\label{sec:review_SF}

In this section, we give a brief review of the 4D Lorentzian spinfoam model with a cosmological constant first introduced in \cite{Han:2021tzw} and recently improved in \cite{Han:2025abc2}. The description here is based on the latter. We will focus on the classical theory and critical points of the spinfoam amplitude, which are necessary ingredients for geometrical reconstruction, and refer readers to \cite{Han:2021tzw,Han:2025abc2} for a more detailed description of how the amplitude is constructed. 

Consider a simplicial 4-complex whose dual graph, also called the spinfoam graph, is such that its edges can be dressed with number $0,\cdots,4$ in a way as a {\it colored graph} in colored tensor model \cite{Gurau:2011aq,Gurau:2011xp}. We say such spinfoam graphs are colorable. An example of colorable spinfoam graph is the melonic spinfoam graph discussed in \cite{Han:2023hbe}. 
The spinfoam amplitude for a 4-complex respects the {\it local amplitude ansatz} and is written as the product of vertex amplitudes, each associated to one 4-simplex in the simplicial decomposition or one vertex in the spinfoam graph, edge amplitudes, each associated to one internal tetrahedron or one edge in the spinfoam graph, and face amplitudes, each associated to one internal triangle or one face in the spinfoam graph.

We first define the vertex amplitude, which is the key to the spinfoam model and encodes the dynamics of LQG. The vertex amplitude can be understood as the constrained $\SL(2,\bC)$ Chern-Simons partition function on $\SG$. 
The Chern-Simons theory has a complex coupling constant 
\be
t=k+is\,,\quad k=8N=\f{3}{2G\hbar\gamma|\Lambda|},\quad N\in\Z_+\,,\quad s=\gamma k\,,
\label{eq:t_k_hbar}
\ee
where $\gamma\in\R$ is the Barbero-Immirzi parameter. The kinematical phase space is the moduli space of $\SL(2,\bC)$ flat connection on the boundary of $\SG$, denoted as
\be
\cP_{\partial(\SG)}=\cM_\Flat(\partial(\SG),\SL(2,\bC))\,.
\ee
It is (complex) 30-dimensional. 
The boundary $\partial(\SG)$ is composed of five 4-holed spheres, each produced by removing the open ball around a vertex of $\Gamma_5$, and ten annuli, each produced by removing the open neighbourhood of an edge of $\Gamma_5$. 
To describe $\cP_{\partial(\SG)}$, we choose a set of symplectic coordinates locally associated to the 4-hole spheres or annuli. These coordinates are packaged in the vector $\vec{\cQ}$ of position variables and the vector $\vec{\cP}$ of conjugate momentum variables with elements
\be\begin{aligned}
\vec{\cQ}&=(2L_{12},2L_{13},2L_{14},2L_{15},2L_{23},2L_{24},2L_{25},2L_{34},2L_{35},2L_{45},M_1,M_2,M_3,M_4,M_5)^\top\,,\\
\vec{\cP}&=(T_{12},T_{13},T_{14},T_{15},T_{23},T_{24},T_{25},T_{34},T_{35},T_{45},P_1,P_2,P_3,P_4,P_5)^\top\,.
\end{aligned}
\label{eq:def_cQ_cP}
\ee
$\vec{\cQ}$ and $\vec{\cP}$ are in fact logarithmic coordinates of $\vec{\fq}\equiv e^{\vec{\cQ}}$ and $\vec{\fp}\equiv e^{\vec{\cP}}$ respectively with a (randomly) chosen branch.
Each element in $\vec{\cQ}$ and $\vec{\cP}$ is a complex variable and they satisfy the Poisson brackets
\be
\{\cQ_I,\cP_J\}=\delta_{IJ}\,,\quad \forall\,I,J=1,\cdots,15\,.
\ee
$(M_a,P_a)$ is a pair of holomorphic (logarithmic) Fock-Gontrarov (FG) coordinates associated to the 4-holed sphere $\cS_a$ ($a=1,\cdots,5$), and $(2L_{ab},T_{ab})$ is a pair of holomorphic (logarithmic) Fenchel-Nielsen (FN) coordinates associated to the annulus $(ab)$ connecting $\cS_a$ and $\cS_b$. The position variable $2L_{ab}$ is called the FN length, and its conjugate momentum $T_{ab}$ is called the FN twist.  

\begin{figure}[h!]
\centering
\includegraphics[width=0.7\textwidth]{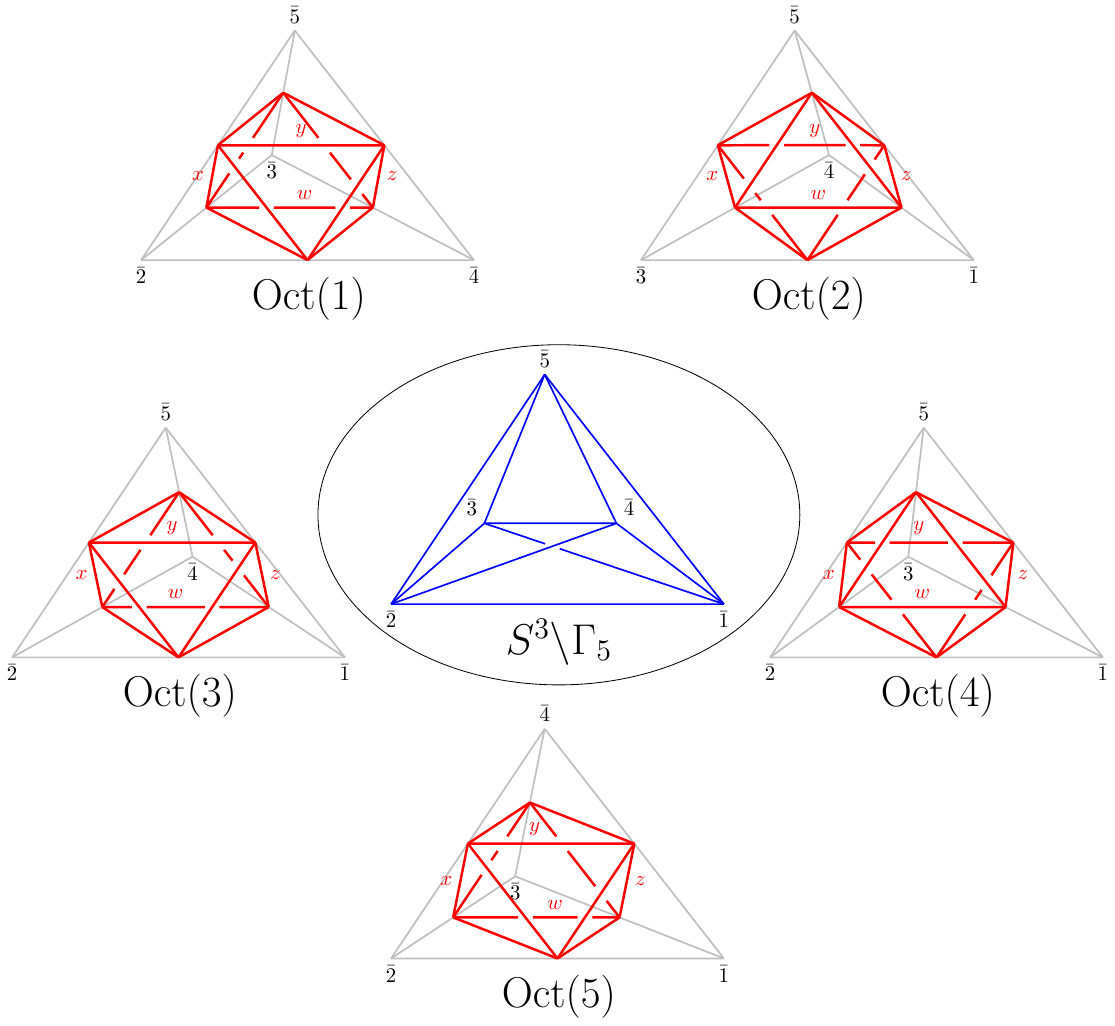}	
\caption{The decomposition of the ideal triangulation $\TSG$ of $\SG$ into 5 ideal octahedra ({\it in red}). Numbers $\bar{1},\bar{2},\bar{3},\bar{4},\bar{5}$ with bars denote the 4-holed spheres on $\partial (S^3\backslash\Gamma_3)$. 
In each ideal octahedron $\Oct(i)$, $x, y, z, w$ ({\it labelled in red}) are chosen to form the equator of the octahedron. }
\label{fig:triangulation_All}
\end{figure}

\begin{figure}[h!]
\centering
\begin{minipage}{0.45\textwidth}
\includegraphics[width=0.5\textwidth]{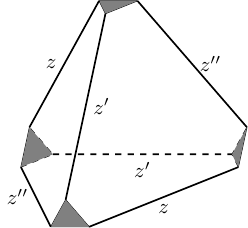}
\subcaption{}
\label{fig:ideal_tetra}
\end{minipage}
\begin{minipage}{0.45\textwidth}
\includegraphics[width=0.6\textwidth]{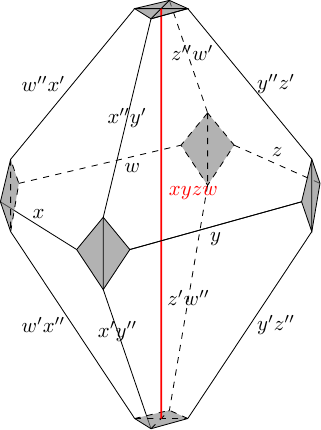}
\subcaption{}
\label{fig:ideal_octa}
\end{minipage}
\caption{{\it (a)} An ideal tetrahedron. {\it (b)} An ideal octahedron composed of 4 ideal tetrahedra. The part of boundaries shaded in gray are called the cusp boundaries and they are located on the boundaries of $\SG$ in $\TSG$. See \eg \cite{Han:2021tzw} for more descriptions. }
\label{fig:ideal_oct_tetra}
\end{figure}

Another way to decompose $\partial(\SG)$ is induced by the so-called {\it ideal triangulation}, denoted as ${\bf T}(\SG)$, of $\SG$, on which the Chern-Simons partition function is well-known \cite{Fock:2003alg,Gaiotto:2009hg}. 
The ideal triangulation of $\SG$, as illustrated in fig.\ref{fig:triangulation_All} is composed of 5 {\it ideal octahedra} denoted as $\Oct(i)$ ($i=1,\cdots,5$), each of which is further composed of 4 {\it ideal tetrahedra} denoted as $\Delta_{x_i},\Delta_{y_i},\Delta_{z_i},\Delta_{w_i}$ with one internal edge. An ideal tetrahedron and an ideal octahedron are illustrated in fig.\ref{fig:ideal_oct_tetra}. 
In an ideal tetrahedron $\Delta$, each of its 6 edges are associated with holomorphic FG coordinates from the set $\{z,z',z''\}$ in the way as in fig.\ref{fig:ideal_tetra} as well as anti-holomorphic FG coordinates from the set $\{\zt,\zt',\zt''\}$ in the same way. These coordinates satisfy 
\be
zz'z''=-1,\quad \zt\zt'\zt''=-1
\ee
which eliminates one of the coordinates from each set, say $z'$ and $\zt'$. Then $(z,z'')$ (\resp $(\zt,\zt'')$) is a pair of holomorphic (\resp anti-holomorphic) symplectic coordinates of the Chern-Simons phase space $\cP_{\partial\Delta}=\cM_\Flat(\partial\Delta,\SL(2,\bC))$ on the boundary $\partial\Delta$ of $\Delta$. 
The logarithmic variables $Z=\ln z,\,Z''=\ln z'',\,\Zt=\ln \zt,\, \Zt''=\ln \zt''$ satisfy the Poisson brackets
\be
\{Z,Z''\}=\{\Zt,\Zt''\}=1\,,\quad \{Z,\Zt\}=\{Z,\Zt''\}=\{Z'',\Zt\}=\{Z'',\Zt''\}=0\,.
\ee
The solution space of the Chern-Simons theory in $\Delta$ is given by the flatness conditions
\be
z''+z^{-1}-1=0\,,\quad \zt''+\zt^{-1}-1=0
\label{eq:flatness}
\ee
which restricts $\cP_{\partial\Delta}$ to its Lagrangian submanifold
\be
\cL_\Delta=\{(z,z'';\zt,\zt'')\in\cP_{\partial\Delta}|z''+z^{-1}-1=0,\zt''+\zt^{-1}-1=0\}\,.
\ee
The quantization of \eqref{eq:flatness} into constraint operator solves for the wave function, or the Chern-Simons partition function on $\Delta$, which turns out to be the quantum dilogarithm function $\Psi_\Delta(z,\zt)$ \cite{Faddeev:1993rs,Dimofte:2011gm}. The exact expression of $\Psi_\Delta(z,\zt)$ is not important in the following analysis, so we omit it here. 

As an ideal octahedron is formed by gluing four ideal tetrahedra ({\it r.f.} fig.\ref{fig:ideal_octa}), the partition function $\cZ_{\Oct}$ is a product of four quantum dilogarithm functions. Note that there is one internal edge, a flatness condition has to be imposed on the FG coordinates $x,y,z,w$ and $\xt,\yt,\zt,\wt$ from the four ideal tetrahedra dressing the internal edge to enforce flat connection in the bulk, which reads
\be
xyzw=1\,,\quad \xt\yt\zt\wt=1\,.
\ee
It eliminates the variables from one of the ideal tetrahedra, say $w,\wt$ from $\Delta_{w}$. The phase space $\cP_{\partial\Oct}$ on $\partial\Oct$ is spanned by position variables $x,y,z$ and momentum variables $p_x=x''-w'',p_y=y''-w'',p_z=z''-w''$.

Since all the edges of ideal tetrahedra in $\TSG$ are on the boundary ({\it r.f.} fig.\ref{fig:triangulation_All}), no additional constraints need to be added. This means the phase space $\cP_{\partial(\SG)}$ on $\partial(\SG)$ is simply the direct product of five copies of $\cP_{\partial\Oct(i)}$. Package the logarithmic phase space coordinates into vectors and their anti-holomorphic counterparts:
\be\begin{aligned}
&\vec{\Phi}\equiv \ln \vec{\phi}=(\{X_i,Y_i,Z_i\}_{i=1}^{5})^\top\,,\quad
\vec{\Pi}\equiv \ln \vec{\pi}=(\{P_{X_i},P_{Y_i},P_{Z_i}\}_{i=1}^{5})^\top\,,\quad\\
&\vec{\widetilde{\Phi}}\equiv \ln \vec{\tilde{\phi}}=(\{\Xt_i,\Yt_i,\Zt_i\}_{i=1}^{5})^\top\,,\quad
\vec{\widetilde{\Pi}}\equiv \ln \vec{\tilde{\pi}}=(\{P_{\Xt_i},P_{\Yt_i},P_{\Zt_i}\}_{i=1}^{5})^\top\,,
\end{aligned}
\label{eq:phi_pi}
\ee
where $\fZ_i=\ln \fz_i,\,P_{\fZ_i}=\ln p_{\fz_i}$ with $\fz_i=x_i,y_i,z_i,\xt_i,\yt_i,\zt_i$ and $\fZ_i=X_i,Y_i,Z_i,\Xt_i,\Yt_i,\Zt_i$. The elements of $\vec{\Phi},\vec{\Pi},\vec{\widetilde{\Phi}},\vec{\widetilde{\Pi}}$ satisfy the Poisson bracket 
\be
\{\Phi_I,\Pi_J\}=\delta_{IJ}\,,\quad 
\{\Phi_I,\widetilde{\Phi}_J\}=\{\Phi_I,\widetilde{\Pi}_J\}=\{\Pi_I,\widetilde{\Phi}_J\}=\{\Pi_I,\widetilde{\Pi}_J\}=0\,,\quad
\forall\, I,J=1,\cdots, 15\,.
\ee

The partition function on $\SG$ is simply the product of five $\cZ_{\Oct}$'s:
\be
\cZ_\times(\{x_i,y_i,z_i;\xt_i,\yt_i,\zt_i\}_{i=1}^5) = \prod_{i=1}^{5}\cZ_{\Oct}(x_i,y_i,z_i;\xt_i,\yt_i,\zt_i)\,.\,,
\label{eq:Z_times}
\ee
where 
\be
\cZ_{\Oct}(x,y,z;\xt,\yt,\zt) = \Psi_\Delta(x,\xt)\Psi_\Delta(y,\yt)\Psi_\Delta(z,\zt)\Psi_\Delta(\f{1}{xyz},\f{1}{\xt\yt\zt})\,.
\ee

However, unlike $\{\cQ_I,\cP_I\}_{I=1}^{15}$ defined in \eqref{eq:def_cQ_cP}, the coordinates $\{X_i,Y_i,Z_i\}_{i=1}^5$ are not localized coordinates on $\partial(\SG)$, rendering it difficult to give them geometrical interpretations. For this reason, one would like to express the Chern-Simons partition function on $\SG$ in terms of $\vec{\cQ}$ and its anti-holomorphic counterpart $\vec{\widetilde{\cQ}}$, which corresponds to a series of Weil transformations on $\cZ_\times$. The Weil transformations reflect the symplectic transformation from coordinates $(\vec{\Phi},\vec{\Pi})$ and $(\vec{\widetilde{\Phi}},\vec{\widetilde{\Pi}}\})$ to $(\vec{\cQ},\vec{\cP})$ and $(\vec{\widetilde{\cQ}},\vec{\widetilde{\cP}})$, which is summarized as follows.
\be
\mat{c}
{\vec{\cQ} \\\vec{\cP}}
=\mat{cc}{\bA&\bB\\{\bf C}&{\bf D}}\mat{c}{\vec{\Phi}\\\vec{\Pi}}+i\pi \mat{c}{\vec{t}_\alpha\\\vec{t}_\beta}\,,\quad
\mat{c}
{\vec{\widetilde{\cQ}} \\\vec{\widetilde{\cP}}}
=\mat{cc}{\bA&\bB\\{\bf C}&{\bf D}}\mat{c}{\vec{\widetilde{\Phi}}\\\vec{\widetilde{\Pi}}}-i\pi \mat{c}{\vec{t}_\alpha\\\vec{t}_\beta}\,,
\label{eq:symp_transf}
\ee
where $\bA,\bB,{\bf C}$ and ${\bf D}$ matrices are all $15\times 15$ matrices with half-integer entries and $\vec{t}_\alpha$ and $\vec{t}_\beta$ are length-$15$ vectors with half-integer entries. We refer to \cite{Han:2025abc2} for their explicit expressions. 

Other than decomposing the boundary of ${\bf T}(\SG)$ into ideal octahedra, one can also decompose it into ideal triangulations of 4-holed sphere $\cS_a$'s, each denoted as ${\bf T}(\cS_a)$, and ideal triangulations of annuli. ${\bf T}(\cS_a)$ is combinatorially the boundary of a tetrahedron with cusp vertices located at the holes (see more detailed description in \eg \cite{Han:2023hbe}). Each of the six edges in ${\bf T}(\cS_a)$ is shared by two ideal octahedra, say $\Oct(i)$ and $\Oct(j)$. It is dressed with an FG coordinates denoted as $\chi^{(a)}_{ij}$. 
Among the coordinates $\vec{\cQ}$ and $\vec{\cP}$ ({\it r.f.} \eqref{eq:def_cQ_cP}), $\{\cQ_I\}_{I=1}^{15}$ and $\{P_a\}_{a=5}$ are related to the set of FG coordinates $\{\chi^{(a)}_{ij}\}_{a,i,j}$ by linear combinations. We give the explicit expressions in \eqref{eq:chi_to_LMP} and \eqref{eq:L} in Appendix \ref{app:symplec_tranf}. This relation will be used for geometrical reconstruction in Section \ref{subsec:FG_on_Sa}. 

As a result, the partition function that is used to define the vertex amplitude is
\begin{multline}
\cZ_{\SG}(\vec{\mu}|\vec{m})\equiv\cZ_{\SG}(\vec{\cQ},\vec{\widetilde{\cQ}}) 
=\frac{8i}{k^{30}}e^{\frac{\pi Q}{k}\vec{\mu}\cdot\vec{t}_{\beta}}e^{\frac{i\pi}{k}\left(\vec{m}\cdot{\bf CA^{-1}}\cdot\vec{m}-\left(\vec{\mu}-i\f{Q}{2}\vec{t}_{\alpha}\right)\cdot{\bf CA^{-1}}\cdot\left(\vec{\mu}-i\f{Q}{2}\vec{t}_{\alpha}\right)\right)}\:\\
\sum_{\vec{n}\in4\left(\mathbb{Z}/2N\mathbb{Z}\right)^{15}}
\int{\rm d}^{15}\nu\,
\sum_{\vec{m}_0\in\left(\mathbb{Z}/k\mathbb{Z}\right)^{15}}\int{\rm d}^{15}\mu_0\:e^{\frac{i\pi}{k}\left(\vec{\nu}\cdot{\bf BA^{\top}}\cdot\vec{\nu}-\vec{n}\cdot{\bf BA^{\top}}\cdot\vec{n}\right)}e^{\frac{2\pi i}{k}\left(\left({\bf A}\vec{\mu}_0-\vec{\mu}+i\f{Q}{2}\vec{t}_{\alpha}\right)\cdot\vec{\nu}+\left(\vec{m}-{\bf A}\vec{m}_0\right)\cdot\vec{n}\right)}\mathcal{Z}_{\times}(\vec{\mu}_0|\vec{m}_0)\,,
\label{eq:SG_partition_final}
\end{multline}
where $\vec{\mu},\vec{\nu},\vec{m},\vec{n}$ comes from the parametrizations of $\vec{\cQ},\vec{\widetilde{\cQ}},\vec{\cP},\vec{\widetilde{\cP}}$ as follows.
\be
\vec{\cQ}=\f{2\pi i}{k}\lb -ib\vec{\mu}-\vec{m} \rb\,, \quad
\vec{\cP}=\f{2\pi i}{k}\lb -ib\vec{\nu}-\vec{n} \rb \,,\quad
\vec{\widetilde{\cQ}}=\f{2\pi i}{k}\lb -ib^{-1}\vec{\mu}+\vec{m} \rb\,, \quad
\vec{\widetilde{\cP}}=\f{2\pi i}{k}\lb -ib^{-1}\vec{\nu}+\vec{n} \rb \,.
\label{eq:P_Q_param}
\ee
In the above parametrizations, $b$ is a phase defined in terms of the Barbero-Immirzi parameter $\gamma$ by
\be
b^2=\frac{1-i \gamma}{1+i \gamma}, \quad \re(b)=:\f{Q}{2}>0, \quad \im(b) \neq 0, \quad|b|=1\,.
\label{eq:b}
\ee
$Q=b+b^{-1}$ is twice the real part of $b$. In the parametrization \eqref{eq:P_Q_param}, $\vec{\mu},\vec{\nu}\in\R^{15}$ are real vectors while $m_I\sim m_I+k,n_I\sim n_I+k$ are periodic classically, leading to discrete spectra $m_I,n_I\in \Z/k\Z$ hence they appear in the sum in the partition function expression \eqref{eq:SG_partition_final}. Similarly to \eqref{eq:P_Q_param}, $\vec{\mu}_0,\vec{m}_0$ in the argument of $\cZ_\times$ in \eqref{eq:SG_partition_final} are the parametrizations of $\vec{\Phi}$ an $\vec{\widetilde{\Phi}}$. Together with their conjugate momenta, they read
\be
\vec{\Phi}=\frac{2\pi i}{k}\lb-ib\, \vec{\mu}_0-\vec{m}_0\rb\,,\quad
\vec{\Pi}=\frac{2\pi i}{k}\lb-ib\, \vec{\nu}_0-\vec{n}_0\rb\,,\quad
\vec{\widetilde{\Phi}}=\frac{2\pi i}{k}\lb-ib^{-1}\, \vec{\mu}_0+\vec{m}_0\rb\,,\quad
\vec{\widetilde{\Pi}}=\frac{2\pi i}{k}\lb-ib^{-1}\, \vec{\nu}_0+\vec{n}_0\rb\,.
\ee
The integration contours in \eqref{eq:SG_partition_final} are shifted from real space to $\R^{15}+i\vec{\alpha}$ with a constant distance $\vec{\alpha}$, called the positive angle, in order to guarantee the boundedness of \eqref{eq:SG_partition_final}. See \cite{Han:2025abc2} for more details. In the semi-classical regime ($\hbar\rightarrow 0$ or $k\rightarrow\infty$, {\it r.f. }\eqref{eq:t_k_hbar}), which is what we focus on in this paper, this shift is invisible so the integration can be approximately taken as along $\R^{15}$. 
Importantly, as there are half-integers in the entries of $\bA,\bB,{\bf C}$ and ${\bf D}$, we only consider $\vec{m}\in4(\Z/2N\Z)^{15}$ in $\cZ_{\SG}(\vec{\mu}|\vec{m})$ for this to be well-defined as the Chern-Simons partition function on $\SG$ \cite{Han:2025abc2}.  

\medskip

To define the vertex amplitude, we impose the {\it simplicity constraints} on $\cZ_{\SG}(\vec{\mu}|\vec{m})$. The result gives a restriction on part (but not all) of the elements in $\vec{\cQ},\vec{\widetilde{\cQ}},\vec{\cP},\vec{\widetilde{\cP}}$ to be coordinates of $\cM_\Flat(\SG,\SU(2))$ in the classical limit. Explicitly, we first restrict
\be
2L_{ab}\equiv \f{2\pi i}{k}\lb -ib\mu_{ab}-m_{ab}\rb=4\pi i j_{ab}=-2\widetilde{L}_{ab}
\quad\Longleftrightarrow\quad
\mu_{ab}=0\,,\quad m_{ab}=2j_{ab}\,,\quad
j_{ab}\in \{0,2,\cdots, 4N-2\}\,.
\label{eq:1st_simplicity}
\ee
The spin $j_{ab}$ encodes the area spectrum $\fa_{ab}$ of the triangle dual to the annulus by
\be
\f{|\Lambda|}{3}\fa_{ab}=\begin{cases}
\f{4\pi }{k}j_{ab}\,,\quad & \text{ if } \,j_{ab}\in[0,2N-2]\\
\lb 2\pi-\f{4\pi }{k}j_{ab}\rb \,,\quad & \text{ if } \,j_{ab}\in[2N,4N-2]
\end{cases}\,. 
\label{eq:spin_to_area}
\ee
Secondly, we couple $\cZ_{\SG}(\vec{\mu}|\vec{m})$ with 5 coherent states $\{\Psi_{\hrho_a}(\mu_a|m_a)\}_{a=1}^5$ (see \cite{Han:2025abc2} for the explicit expression), each associated to one 4-holed sphere $\cS_a$ on $\partial(\SG)$ as a function of $M_a=\f{2\pi i}{k}\lb -ib\mu_a-m_a \rb$. 
We will also use the parametrization $P_a=\f{2\pi i}{k}\lb -ib\nu_a-n_a \rb$ in the following.
The coherent state label $\hrho_a\equiv (\zh_a,\xh_a,\yh_a )\in \bC\times [0,2\pi)\times [0,2\pi)$ encodes the shape of the tetrahedron dual to $\cS_a$ with fixed triangle areas $\{\fa_{ac}\}_{c\neq a}$. 

As a result, the vertex amplitude is defined as

\be
\cA_v(\iota):=\sum_{\{m_a\}\in 4(\Z/2N\Z)^5}\int_{\R^5+i\beta_a} \rd^5\mu_a \,
\cZ_{S^3\backslash\Gamma_5}(\{i\alpha_{ab}\}_{a<b}, \{\mu_{a}+i\alpha_a\}\mid\{j_{ab}\}_{a<b}, \{m_{a}\})
\prod\limits_{a=1}^5\Psi_{\hrho_a}(\mu_a|m_a)\,,
\label{eq:vertex_amplitude}
\ee
where $\iota=(\{\alpha_{ab},j_{ab}\}_{a<b}, \{\hrho_a\}_{a=1}^5, \{\alpha_a,\beta_a\}_{a=1}^5)$ with $\{\alpha_{ab}\}_{a<b},\{\alpha_a,\beta_a\}_a$ being positive angles that are omitted at the semi-classical regime. 

Consider a simplicial 4-complex ${\bf T}(M_4)$ as the triangulation of a 4-manifold $M_4$ whose simplicial decomposition consists of $V$ 4-simplices, $E_{\In}$ internal tetrahedra and $F_{\In}$ internal edges and whose spinfoam graph is colorable.
It corresponds to gluing $V$ copies of $\SG$'s by identifying $E_{\In}$ pairs of 4-holed spheres and forming $F_{\In}$ tori from gluing annuli. Restricting to colorable spinfoam graphs means that each pair of glued 4-holed spheres share the same label, \ie $\cS_a$ is only glued to $\cS'_b$ with $b=a$. 
 The amplitude for the 4-complex is composed of $V$ vertex amplitudes $\cA_v$'s, $E_{\In}$ edge amplitudes $\cA_e$'s and $F_{\In}$ face amplitudes $\cA_f$'s. It takes the form 

\be
\cZ_{\vec{\hrho}_\partial}(\vec{\alpha}|\vec{j}_b)
=\sum_{{\rm even }\,j_f=0}^{4N-2}
\int_{\overline{\cM}_{\vec{j}^v_a}}\rd \hrho^{v\in e}_a\int_{\overline{\cM}_{\vec{j}^{v'}_a}}\rd \hrho^{v'\in e}_a 
\left[\prod_{f=1}^{F_{\In}}\mathcal{A}_f(2j_f)\right]
\left[\prod_{e=1}^{E_{\In}}
\cA_e(\hrho_a^{v\in e},\hrho_a^{v'\in e}|\{j_{ac}^{v\in e},j_{ac}^{v'\in e}\}_{c\neq a})\right]
\left[\prod_{v=1}^V\cA_{v}(\vec{\alpha}^v,\vec{j}^v,\vec{\hrho}^v)\right],
\label{eq:spinfoam_amplitude}
\ee
where $v\in e$ denotes that $v$ is at the (source or target) end of $e$, $\vec{\alpha}$ contains all the positive angles, $\vec{\hrho}_\partial$ contains all the coherent state labels on the boundary, the summations in $j_f$ are for all the internal spinfoam faces and the integrations over coherent state labels are for all the internal spinfoam edges. 

The edge amplitude for spinfoam edge $e$ connecting spinfoam vertices $v$ and $v'$ that corresponding to gluing $\cS_a$ and $\cS'_a$ from different $\SG$'s is defined as
\be
\cA_e^{vv'}\lb\hrho_a^{v\in e},\hrho_a^{v'\in e}|\{j_{ac}^{v\in e},j_{ac}^{v'\in e}\}_{c\neq a}\rb 
:= \f{k}{(2\pi)^4}\delta_{\vec{j}_a^{v}-\vec{j}_a^{v'}}
\delta_{\hat{\rho}_a'-\hat{\tilde{\rho}}_a}\delta_{\alpha_a'+\alpha_a} e^{-\f{2\pi (\beta_a+\beta'_a)}{k}}\,,
\label{eq:edge_amplitude}
\ee
where $\hat{\tilde{\rho}}_a=(-\bar{\hat{z}}_a,-\xh_a,\yh_a)$. The integrations $\overline{\cM}_{\vec{j}^v_a}$ and $\overline{\cM}_{\vec{j}^{v'}_a}$ in \eqref{eq:spinfoam_amplitude} are over compact spaces of the curved tetrahedron shapes given triangle areas fixed by spins \cite{Han:2023hbe}. 

The face amplitude for spinfoam face $f$ is defined as the function of spin $j_f$, which is the constrained FN length associated to the torus. It takes the form
\be
\cA_f(2j_f):=[2j_f+1]_\fq^\fp \,,\quad\fp\in \R\,,\quad
j_f=0,2,\cdots,4N-2\,,
\label{eq:face_amplitude}
\ee
where $[n]_\fq:=\f{\fq^n-{\fq}^{-n}}{\fq-\fq^{-1}}\equiv \sin\lb \f{2\pi n}{k} \rb/\sin\lb \f{2\pi}{k}\rb$ is a $\fq$-number with $\fq=e^{2\pi i/k}$ being a root-of-unity depending on the Chern-Simons level $k$. 

Other than viewing $\cZ_{\vec{\hrho}_\partial}(\vec{\alpha}|\vec{j}_b)$ as the amplitude defined from the local amplitude ansatz, it can also be viewed as the constrained partition function on a graph-complement 3-manifold $M_3\backslash\Gamma$, where $M_3=\partial M_4$ and $\Gamma$ is the dual graph of the triangulation ${\bf T}(M_3)$ of $M_3$. Before imposing the constraints, the partition function is the function of the position variables of the phase space $\cM_\Flat(\partial(M_3\backslash\Gamma),\SL(2,\bC))$. 

At large $k$ regime, \eqref{eq:spinfoam_amplitude} takes the form of the path integral of an oscillatory action, rendering the applicability of stationary phase analysis to solve for the critical point. The critical solutions are as follows. Firstly, simultaneously for all 4-simplices in the simplicial decomposition of ${\bf T}(M_4)$, the symplectic transformation \eqref{eq:symp_transf} is recovered at the critical point. Secondly, the parameters $\vec{\mu}_0,\vec{\nu}_0,\vec{m}_0,\vec{n}_0$, hence the coordinates $\vec{\Phi},\vec{\widetilde{\Phi}},\vec{\Pi},\vec{\widetilde{\Pi}}$ satisfy the flatness conditions similar to \eqref{eq:flatness}. This means the critical points of the spinfoam amplitude to are points of $\cM_\Flat(M_3\backslash \Gamma,\SL(2,\bC))$.
In addition, for each boundary tetrahedron, or boundary 4-holed sphere on $\partial(M_3\backslash\Gamma)$, the peaks of $M_a,P_a$ are determined by the coherent state label $\hrho_a$ hence the tetrahedron shape.
\be
\mu_{a}=\frac{k}{\sqrt{2}\pi}{\rm Re}(z_{a})\,,\quad
\nu_{a}=-\frac{k}{\sqrt{2}\pi}{\rm Im}(z_{a})\,,\quad
m_{a}=\frac{k}{2\pi}\hat{x}_{a}\,,\quad
n_{a}=-\frac{k}{2\pi}\hat{y}_{a}\,.
\ee
This tricical point solution means that the Chern-Simons phase space $\cM_\Flat(\partial\cS_a,\SL(2,\bC))$ is restricted to the tetrahdedron phase space isomorphic to $\cM_{\Flat}(\partial\cS_a,\SU(2))$. 
Lastly, the critical point corresponding to the spin $j_f$ for internal triangle $\triangle_f$ gives a vanishing deficit angle hinged by $\triangle_f$. We refer to \cite{Han:2025abc2} for a full derivation of these critical solutions. 

\section{Geometrical reconstruction of spinfoam critical points}
\label{sec:geo_reconstruct}

Now that we have specified the critical points of the spinfoam amplitude to be points of $\cM_\Flat(M_3\backslash \Gamma,\SL(2,\bC))$ where $M_3\backslash\Gamma$ can be formed by gluing $\SG$'s and that the generators of $\cM_\Flat(\SG,\SL(2,\bC))$ can be represented by holonomies in terms of the geometrical data of a curved 4-simplex ({\it r.f.} Section \ref{sec:flat_connection_2_geometry}). It remains to encode this geometrical information to the coordinates $(\vec{\cQ},\vec{\cP})$.
The bridge to connect $\{\vec{\cQ},\vec{\cP}\}$ and the curved 4-simplex geometry is provided by the {\it framing flags} on $\cL_{\SG}$ as they can be constructed using geometrical variables and can be used to formulate the FG coordinates. 
A framing flag is a choice of flat section $s=\lb s^0,s^1 \rb^\top\in\bC^2$ in an associated $\bC\bP^1$ bundle over every cusp boundary satisfying $\rd s=As$. We first briefly summarize how the FG coordinates are defined with framing flags. See \eg \cite{Dimofte:2011gm,Dimofte:2014zga,Gaiotto:2009hg} for more details. 

\begin{figure}[h!]
\centering
\includegraphics[width=0.2\textwidth]{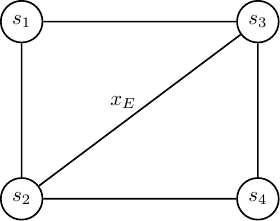}
\caption{A quadrilateral in a 2D ideal triangulation to define FG coordinate $x_E$ in terms of the framing flags $\{s_i\}_{i=1,\cdots,4}$ by \eqref{eq:FG_from_flag}. }
\label{fig:flag}
\end{figure}

Given a 2D ideal triangulation of the boundary $\partial(\cM\backslash\Gamma)$ of a graph $\Gamma$-complement 3-manifold $\cM\backslash\Gamma$, a framing flag is chosen at each cusp boundary. Each edge $E$ of the ideal triangulation can be organized to be the diagonal of a quadrilateral as shown in fig.\ref{fig:flag}. 
Parallel transport the framing flags to a common point inside the quadrilateral and denote them as $s_1,\cdots,s_4$ respectively. Then the FG coordinate $x_E$ on $E$ is then defined as
\be
x_E=\frac{\left\langle s_1 \wedge s_2\right\rangle\left\langle s_3 \wedge s_4\right\rangle}{\left\langle s_1 \wedge s_3\right\rangle\left\langle s_2 \wedge s_4\right\rangle}\,,\quad
\left\langle s_i \wedge s_j\right\rangle:=s_i^0 s_j^1-s_i^1s_j^0\,.
\label{eq:FG_from_flag}
\ee
This relation is indeed invariant under the complex rescaling of any $s_i$ and the inner bracket is $\SL(2,\bC)$-invariant. 

In this section, we will first focus on the geometrical reconstruction for the FG coordinates on a 4-holed sphere $\cS_a$ in Section \ref{subsec:FG_on_Sa} then move to the full set of coordinates on the whole 3-manifold $\SG$ defined in a different way in Section \ref{subsec:coord_on_SG}.

\subsection{Fock-Gontrarov coordinates on one 4-holed sphere}
\label{subsec:FG_on_Sa}

Let us view fig.\ref{fig:flag} as part of the ideal triangulation of $\cS_a$ with cusp boundaries located at the holes. 
Then the framing flag $s_i$ parallel transported from hole $i$ of $\cS_a$ connected to, say, hole $j$ of $\cS_b$, to a common point $\fb_a$ in $\cS_a$ is the eigenvector of holonomy $\Ht_{ab}$. 
When the simplicity constraints are imposed, $\Ht_{ab}$ is conjugate to $O_{ab}\in\SU(2)$ as in \eqref{eq:Ht_to_O} and the eigenvectors are the spinors $|\xi_{ab}\ra$ and $|\xi_{ab}]$.
This means that the role of framing flags can be played by the spinors when the simplicity constraints are imposed. The spinors, in turn, can be constructed by the normal vectors of the triangles in the local frame of the tetrahedron according to \eqref{eq:xi_from_n}. This gives a way to encode the geometry of a curved tetrahedron as part of the curved 4-simplex in coordinates of $\cM_{\Flat}(\cS_a,\SU(2))$. 

\begin{figure}[h!]
\centering
\includegraphics[width=0.5\textwidth]{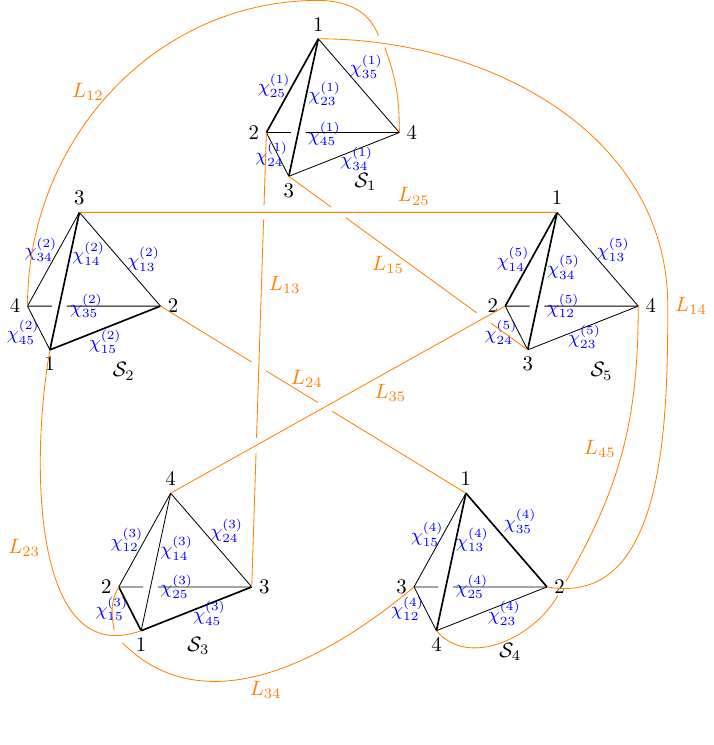}
\caption{Identifying the holes of 4-holed spheres for different $\cS_a$'s. The tetrahedra are the ideal triangulations ${\bf T}(\cS_a)$'s with cusp boundaries shrunk to points dressed with numbers $1,2,3,4$. Each edge of ${\bf T}(\cS_a)$ is dressed with a FG coordinate $\cX_{ij}^{(a)}$. Curves in orange present the annuli, each dressed with an FN length $L_{ab}$.}
\label{fig:identify}
\end{figure}
\begin{figure}[h!]
\centering
\includegraphics[width=0.4\textwidth]{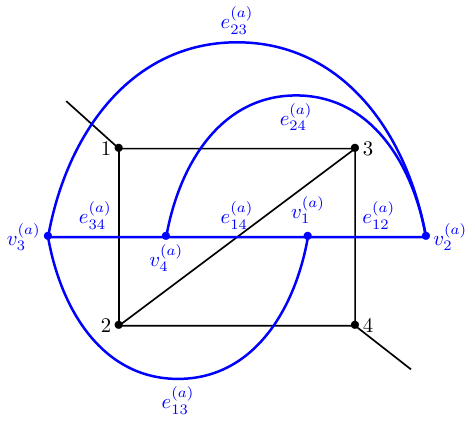}
\caption{The triangulation $T_a$ ({\it in blue}) of $\cS_a$. The cusps of ${\bf T}(\cS_a)$ are shrunk to points labeled by number $i=1,\cdots,4$. Each cusp $i$ is in the triangle of $T_a$ bounded by three nodes $v^{(a)}_j$'s with $j\neq i$.}
\label{fig:S2_T2}
\end{figure}

To be explicit, we first fix some notations. 
We label the holes of each 4-holed sphere $\cS_a$, or equivalently the cusp boundaries of its ideal triangulation $\TcS{a}$, by numbers 1,2,3,4 and fix the gluing of holes in the way illustrated in fig.\ref{fig:identify}. 
The triangulation $T_a$ of $\cS_a$ is the dual graph of $\TcS{a}$. 
Denote the node of $T_a$ opposite to cusp $i$ of $\TcS{a}$ by $v^{(a)}_i$ and the link of $T_a$ connecting $v^{(a)}_i$ and $v^{(a)}_j$ as $e_{ij}^{(a)}$, as illustrated in fig.\ref{fig:S2_T2}. We will also use the same convention for spinors as in the previous section. 

To proceed, we first show that four spinors parallel transported from 4 holes of $\cS_a$ to a common point and four spins associated to the holes do reproduce the FG and FN coordinates with simplicity constraints imposed. Referring to fig.\ref{fig:flag}, denote by $\xi_i$ (\resp $\zeta_i$) the spinor parallel transported from hole $i$ at a point $\fb_a$ (\resp a point $\fb'_a$) living at the middle-point of the edge connecting holes 2 and 3 (\resp holes 1 and 4). Also denote the spin associated to hole $i$ by $j_i$ and the holonomy around hole $i$ based at $\fb_a$ (\resp at $\fb'_a$) as $O_i$ (\resp $O'_i$). Then the FG coordinate $x_{ij}$ on edge connecting holes $i$ and $j$ are explicitly, 
\be
\begin{aligned}
x_{23}&=\f{[\xi_1|\xi_2\ra[\xi_3|\xi_4\ra}{[\xi_1|\xi_3\ra[\xi_2|\xi_4\ra}=
\f{[\zeta_1|\zeta_2\ra[\zeta_3|O'_3\zeta_4\ra}{[\zeta_1|\zeta_3\ra[\zeta_2|O'_3\zeta_4\ra}\,,\\
x_{12}&=\f{[\xi_3|\xi_1\ra[\xi_2|O^{-1}_2\xi_4\ra}{[\xi_3|\xi_2\ra[\xi_1|O_2^{-1}\xi_4\ra}=
\f{[\zeta_4|\zeta_2\ra[\zeta_1|O'^{-1}_1\zeta_3\ra}{[\zeta_4|\zeta_1\ra[\zeta_2|O'^{-1}_1\zeta_3\ra}\,,\\
x_{24}&=\f{[\xi_3|\xi_2\ra[\xi_4|O_2\xi_1\ra}{[\xi_3|\xi_4\ra[\xi_2|O_2\xi_1\ra}=
\f{[\zeta_1|\zeta_4\ra[\zeta_2|{O'_4}\zeta_3\ra}{[\zeta_1|\zeta_2\ra[\zeta_4|{O'_4}\zeta_3\ra}\,,\\
x_{34}&=\f{[\xi_2|\xi_4\ra[\xi_3|O_3^{-1}\xi_1\ra}{[\xi_2|\xi_3\ra[\xi_4|O_3^{-1}\xi_1\ra}=
\f{[\zeta_1|\zeta_3\ra[\zeta_4|{O'}^{-1}_4\zeta_2\ra}{[\zeta_1|\zeta_4\ra[\zeta_3|{O'}^{-1}_4\zeta_2\ra}\,,\\
x_{14}&=\f{[\xi_2|\xi_1\ra[\xi_4|O_4\xi_3\ra}{[\xi_2|\xi_4\ra[\xi_1|O_4\xi_3\ra}=
\f{[\zeta_3|\zeta_4\ra[\zeta_1|\zeta_2\ra}{[\zeta_3|\zeta_1\ra[\zeta_4|\zeta_2\ra}\,,\\
x_{13}&=\f{[\xi_2|\xi_3\ra[\xi_1|O_3\xi_4\ra}{[\xi_2|\xi_1\ra[\xi_3|O_3\xi_4\ra}=
\f{[\zeta_4|\zeta_1\ra[\zeta_3|{O'_1}\zeta_2\ra}{[\zeta_4|\zeta_3\ra[\zeta_1|{O'_1}\zeta_2\ra}\,.
\end{aligned}
\label{eq:FG_spheres}
\ee 
The holonomies $\{O_i\}_i$ or $\{O'_i\}_i$ are used to parallel transport spinors, the appearance of which in the above expressions is because the FG coordinates have to be defined in terms of framing flags {\it inside} the quadrilateral through \eqref{eq:FG_from_flag}. 
These holonomies admit the factorization as in \eqref{eq:Oab_decomp}. That is,
\be\begin{aligned}
O_i &=M(\xi_i)\diag(e^{-2\pi i\f{j_i}{k}},e^{2\pi i\f{j_i}{k}})M(\xi_i)^{-1}\,,\\
O'_i&=M(\zeta_i)\diag(e^{-2\pi i\f{j_i}{k}},e^{2\pi i\f{j_i}{k}})M(\zeta_i)^{-1}\,.
\end{aligned}
\label{eq:O_factorize}
\ee
Plugging this factorization into \eqref{eq:FG_spheres}, one can verify that (see an example in Appendix \ref{app:FG_calculation})
\be\ba{ll}
x_{12}x_{23}x_{24}=-e^{-4\pi i\f{j_2}{k}}\,,\quad& 
x_{13}x_{23}x_{34}=-e^{-4\pi i\f{j_3}{k}}\,,\\[0.15cm]
x_{12}x_{13}x_{14}=-e^{-4\pi i\f{j_1}{k}}\,,\quad&
x_{14}x_{24}x_{34}=-e^{-4\pi i\f{j_4}{k}}\,,
\label{eq:chi_to_2L}
\ea\ee
where the first line is obtained using the expressions in terms of $\xi_i$'s in \eqref{eq:FG_spheres} while the second line is obtained using the expressions in terms of $\zeta_i$'s. 
The {\it r.h.s.} of all equalities in \eqref{eq:chi_to_2L} are nothing but the (exponentiated) FN lengths after imposing the first-class simplicity constraints \eqref{eq:1st_simplicity}. 

The FG coordinates $M_a$ and $P_a$ can also be calculated from 4 FN lengths $\{L_{ab}\}_{b}$ and two selected FG coordinates $x_{ij}$'s obtained geometrically in \eqref{eq:FG_spheres}. We denote the two chosen FG coordinates as $\cX_a$ and $\cY_a$ and fix them to be\footnote{$\cX_a$ and $\cY_a$ are, in fact, the FG coordinates used in the spinfoam model in the original model \cite{Han:2021tzw} (up to sign and constant).} 
\be\begin{aligned}
&\mathcal{X}_1=\chi_{25}^{(1)}\,,\quad  \mathcal{X}_2=\chi_{15}^{(2)}\,,\quad   \mathcal{X}_3=\chi_{15}^{(3)} \,,\quad 
\mathcal{X}_4=\chi_{15}^{(4)}\,,\quad   \mathcal{X}_5=\chi_{14}^{(5)} \,,\\
& \mathcal{Y}_1=\chi_{23}^{(1)} \,,\quad \mathcal{Y}_2=\chi_{14}^{(2)}\,,\quad \mathcal{Y}_3=\chi_{45}^{(3)}\,,\quad
 \mathcal{Y}_4=\chi_{35}^{(4)}\,,\quad \mathcal{Y}_5=\chi_{34}^{(5)}\,.
\end{aligned}\ee
Then $M_a$ and $P_a$ read
\be\begin{aligned}
&M_1= -L_{12}+L_{13}+2 L_{14}-\cX_1-\cY_1+3 i \pi\,,\quad
&&P_1= L_{12}-L_{14}+\cY_1\,,\\
&M_2= L_{23}+L_{25}-\cY_2+3 i \pi \,,\quad
&&P_2= L_{24}-L_{25}-\cX_2\,,\\
&M_3= -L_{13}-L_{34}+\cX_3\,,\quad
&&P_3= -L_{13}-L_{23}-\cY_3+3 i \pi \,\\
&M_4= L_{24}+L_{45}+\cY_4\,,\quad
&&P_4= -2 L_{24}-L_{34}-L_{45}-\cX_4-\cY_4+3 i \pi \,,\\
&M_5= L_{25}-L_{45}+\cX_5+\cY_5\,,\quad
&&P_5= -L_{15}+L_{25}+\cX_5\,.
\end{aligned}
\label{eq:M_P}
\ee 

The above construction for one 4-holed sphere can be generalized to all the 4-holed spheres on the boundary of $\SG$ given geometrical data -- triangle areas and normals. Namely, spinors on $\cS_a$ can be obtained from the normals $\hat{n}_{ab}$ of triangles on the local frame of $T_a$ from \eqref{eq:xi_from_n}, and these spinors together with triangle areas, carrying discrete values $\fa_{ab}$ ({\it r.f.} \eqref{eq:spin_to_area}), describe the holonomies around single holes of $\cS_a$ according to \eqref{eq:O_factorize}, which in turn give a full set of FG coordinates on ${\bf T}(\cS_a)$ according to \eqref{eq:FG_spheres}. 
Note that, if hole $i$ of $\cS_a$ is connected to hole $j$ of $\cS_b$, the corresponding spinors $\xi_{ab}\equiv \xi_i^{(a)}$ and $\xi_{ba}\equiv \xi_j^{(b)}$ are related to by the parallel transport \eqref{eq:xi_G_relation}. 

\begin{figure}[h!]
\centering
\includegraphics[width=0.5\textwidth]{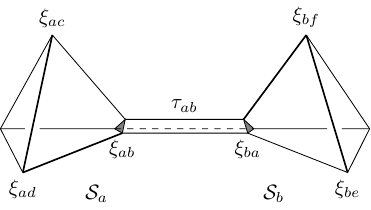}
\caption{The relative positions of the cusps for $\cS_a$ and $\cS_b$ where the associated spinors used to define $\chi_{ab}(\xi)$ \eqref{eq:chi_ab}. The cusps not on the annulus $(ab)$ have been shrunk to points. $\xi_{a*}$'s are defined on the base point $\fb_a$ on $\cS_a$ while $\xi_{b*}$'s are defined on the base point $\fb_b$ on $\cS_b$. }
\label{fig:tau_ab}
\end{figure}
Nevertheless, to get all the elements of $(\vec{\cQ},\vec{\cP})$, one remains to geometrical reconstruct the FN twists $\{\tau_{ab}\equiv e^{T_{ab}}\}_{a<b}$. Each $\tau_{ab}$ is given by the spinors on both $\cS_{a}$ and $\cS_b$ as well as the dihedral angle between the tetrahdera $\tetra_a$ and $\tetra_b$ hinged by the triangle $\triangle_{ab}$ shared by the two tetrahedra \cite{Han:2021tzw} (see also Appendix B of \cite{Han:2024reo}):
\be
\tau_{ab}= e^{-\f12 \nu\sgn(V_4)\Theta_{ab}+i\theta_{ab}}\sqrt{\chi_{ab}(\xi)}\,,
\label{eq:tau_chi}
\ee
where 
\be
\chi_{ab}(\xi)=-\f{[\xi_{be}| \xi_{bf} \ra}{[ \xi_{be}| \xi_{ba} \ra[ \xi_{bf}| \xi_{ba} \ra}
\f{[ \xi_{ac}| \xi_{ab}\ra[ \xi_{ad}| \xi_{ab} \ra}{[ \xi_{ac}| \xi_{ad}\ra}
\label{eq:chi_ab}
\ee
 depends on another two spinors $\xi_{ac}$ and $\xi_{ad}$ on another two cusps (not connected to $\cS_b$) of $\cS_a$ based at $\fb_a$ and another two spinors $\xi_{be}$ and $\xi_{bf}$ another two cusps of $\cS_b$ based at $\fb_b$. Their relative positions are illustrated in fig.\ref{fig:tau_ab}. $\theta_{ab}$ in \eqref{eq:tau_chi} is a non-geometrical parameter that depends on the boundary condition. When the FN twist is associated to an internal triangle, $\theta_{ab}=0$ when the 4-manifold under consideration is globally time-oriented \cite{Han:2015gma,Han:2024reo}.

\subsection{Coordinate reconstruction on $\SG$}
\label{subsec:coord_on_SG}

Another way to reconstruct the Chern-Simons phase space coordinates is to first construct the coordinates $(\vec{\Phi},\vec{\Pi})$ used to define $\cZ_\times$ \eqref{eq:Z_times}, which are the coordinates on ideal octahedra, then use \eqref{eq:symp_transf} to symplectic transform to $(\vec{\cQ},\vec{\cP})$. Moreover, since there are only 10 cusps on $\partial(\SG)$, the collection of all spinors calculated from the local frame of each of the five tetrahedra contributes some redundant data. 
In this subsection, we will fix 10 spinors out of this collection to reconstruct the elements of $(\vec{\Phi},\vec{\Pi})$ used to define $\cZ_\times$.

\begin{figure}
\centering
\includegraphics{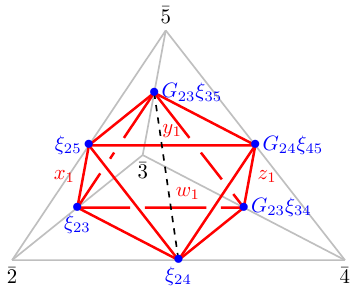}
\caption{Spinors based at $\fb$ on the cusps of $\Oct(1)$.}
\label{fig:oct1}
\end{figure}
Let us fix the single base point $\fb$ of $\SG$ to be the base point $\fb_2$ on $\cS_2$, \ie on the edge connecting cusp 2 and 3 ({\it r.f.} fig.\ref{fig:identify}). 
Then we need to parallel transport the spinors $\{\xi_{ab}\}$ defined at $\fb_a$ for $a\neq 2$ to $\fb$ using the frame-changing holonomy $G_{2a}$ defined in \eqref{eq:G_factorize}. 
Firstly, we focus on the ideal tetrahedron $\Delta_x(1)$ in the ideal octahedron $\Oct(1)$ dressed by FG coordinates $x_1,x'_1$ and $x''_1$ ({\it r.f.} fig.\ref{fig:ideal_octa}). As can be seen from fig.\ref{fig:oct1}, three out of four cusp boundaries of $\Delta_x(1)$ are on the annuli connected to holes of $\cS_2$ while the remaining one is on the annulus connecting $\cS_3$ and $\cS_5$. We define $\xi_{35}$ at $\fb_3$ then use $G_{23}$ that passes along annulus $(23)$ to parallel transport it to $\fb$. Then $x_1$ is expressed in terms of the framing flags $s_1=G_{23}\xi_{35},\,s_2=\xi_{23},\,s_3=\xi_{25},\,s_4=\xi_{24}$
\be
x_1=\frac{\langle G_{23}\xi_{35}\wedge\xi_{23}\rangle\langle\xi_{25}\wedge \xi_{24}\rangle}{\langle G_{23}\xi_{35}\wedge\xi_{25}\rangle\langle\xi_{23}\wedge  \xi_{24}\rangle}\,.
\label{eq:def_x}
\ee
Similarly,
\begin{subequations}
\begin{align}
x'_1&=\frac{\langle\xi_{23}\wedge\xi_{24}\rangle\langle\xi_{25}\wedge G_{23}\xi_{35}\rangle}{\langle\xi_{23}\wedge\xi_{25}\rangle\langle\xi_{24}\wedge G_{23}\xi_{35}\rangle},
\label{eq:def_xp}\\
x''_1&=\frac{\langle\xi_{23}\wedge\xi_{25}\rangle\langle G_{23}\xi_{35}\wedge\xi_{24}\rangle}{\langle\xi_{23}\wedge G_{23}\xi_{35}\rangle\langle\xi_{25}\wedge\xi_{24}\rangle}\,.
\label{eq:def_xpp}
\end{align}
\label{eq:def_xp_xpp1}
\end{subequations}
Using the identity for any three framing flags $s_1,s_2,s_3$:
\be
s_{1}\langle s_{2}\wedge s_{3}\rangle+s_{2}\langle s_{3}\wedge s_{1}\rangle+s_{3}\langle s_{1}\wedge s_{2}\rangle=0\,,
\ee
one can immediately prove that the definitions \eqref{eq:def_x}--\eqref{eq:def_xpp} satisfy the defining relation for $\cL_{\Delta}$, \ie
\be
x_1x'_1x''_1=-1\,,\quad x''_1+x_1^{-1}=1\,.
\ee
The same calculation follows for other ideal tetrahedra as long as the framing flags on the cusp boundaries are fixed. See Appendix \ref{app:FG_on_Oct}. In total, 10 spinors need to be given, one associated to the cusp located on a link of $\Gamma_5$. Keeping in mind that the {\it only} base point of $\SG$ is $\fb_2$, we use $G_{2a}$ to parallel transport the spinor $\xi_{ab}$ to $\fb_2$ when the link $e_{ab}$ is not connected to node $2$. For $a,b\neq2$, we choose the spinor $\xi_{ab}$ with $a<b$ on $\cS_a$ (instead of $\xi_{ba}$ on $\cS_b$). This fixes all the spinors we use:
\be
\ba{llll}
e_{12}: \xi_{21}& 
e_{13}: G_{21}\xi_{13}&
e_{14}: G_{21}\xi_{14}&
e_{15}: G_{21}\xi_{15}\\[0.15cm]
e_{23}: \xi_{23}&
e_{24}: \xi_{24}&
e_{25}: \xi_{25}\\[0.15cm]
e_{34}: G_{23}\xi_{34}&
e_{35}: G_{23}\xi_{35}\\[0.15cm]
e_{45}: G_{24}\xi_{45}
\ea\,.
\label{eq:all_spinors}
\ee
It is straightforward to check that 
\be
x_iy_iz_iw_i=1\,,\quad \forall\,i=1,\cdots,5\,.
\ee
This allows us to eliminate $w_i$ and define
\be
p_{\fz_i}=\fz_i''-w_i''\,,\quad \fz_i=x_i,y_i,z_i\,,
\ee
which is indeed the conjugate momenta of $\fz_i$ as symplectic coordinates of $\cL_{\Oct(i)}$ as the algebraic curve equation 
\be
p_{\fz_i}-x_iy_iz_i+\fz_i^{-1}=0
\ee
is satisfied. We give the explicit formulas \eqref{eq:FG_Oct1} for all the FG coordinates on $\Oct(1)$ in Appendix \ref{app:FG_on_Oct}. 
The vectors of coordinates $\vec{\phi}:=(x_i,y_i,z_i)_{i=1,\cdots,5}^\top$ and $\vec{\pi}:=(p_{x_i},p_{y_i},p_{z_i})_{i=1,\cdots,5}^\top$ are the exponential coordinates of $\vec{\Phi}$ and $\vec{\Pi}$ respectively({\it r.f.} \eqref{eq:phi_pi}). They enter the ``machine'' of symplectic transformation and define the exponentiated FN coordinates on annuli and the exponentiated FG coordinates on 4-holed spheres:
\be\left|\ba{l}
\fq_I=(-1)^{t_{\alpha,I}}\prod\limits_{J,K=1}^{15}\phi_J^{\bA_{IJ}}\pi_K^{\bB_{IK}}\equiv e^{\cQ_I}\\[0.15cm]
\fp_I=(-1)^{t_{\beta,I}}\prod\limits_{J,K=1}^{15}\phi_J^{{\bf C}_{IJ}}\pi_K^{{\bf D}_{IK}}\equiv e^{\cP_I}\ea\right.,\quad
I=1,\cdots,15\,.
\label{eq:q_p_from_phi_pi}
\ee
The parameters of $\vec{\cQ}$ and $\vec{\cP}$ then enters the spinfoam amplitude formula \eqref{eq:vertex_amplitude}. The FN lengths $\{\ell^2_{ab}\equiv\fq_I\}$ with $I=1,\cdots,10$ are those that satisfy the first-class simplicity constraints, \ie $\ell^2_{ab}=e^{-4\pi i j_{ab}/k}$.

\subsection{Compatability test}
\label{subsec:test}

Let us also verify that the coordinate reconstruction for $(\vec{\cQ},\vec{\cP})$ in Section \ref{subsec:FG_on_Sa} and that for $(\vec{\Phi},\vec{\Pi})$ in Section \ref{subsec:coord_on_SG} are compatible. That is, the FG coordinates $\{\chi^{(a)}_{bc}\}$ can be computed from the octahedron coordinates. We examine it with an example: $\chi^{(2)}_{14}=X_1+X_4$, and the same computation can be repeated similarly to check for all the FG coordinates. According to fig.\ref{fig:identify}, $\chi^{(2)}_{14}$ is associated to the edge $e_{13}^{(2)}$ on $\cS_2$ connecting cusps 1 and 3. 
Let $s_{ab}$ be framing flag at the cusp of $\cS_a$ connected to annulus $(ab)$ parallel transported to some point $\fb_0$ on $e_{13}^{(2)}$. 

To do the calculation most easily, we first parallel transport the base point from $\fb$ to $\fb_0$ with holonomy $h_{0\fb}$. Then
\be\begin{aligned}
x_1&=e^{X_1}=\frac{\langle h_{0\fb}G_{23}\xi_{35}\wedge h_{0\fb}\xi_{23}\rangle\langle h_{0\fb}\xi_{25}\wedge  h_{0\fb}\xi_{24}\rangle}{\langle  h_{0\fb}G_{23}\xi_{35}\wedge h_{0\fb}\xi_{25}\rangle\langle h_{0\fb}\xi_{23}\wedge  h_{0\fb} \xi_{24}\rangle}
=\f{\la s_{35} \w s_{23}\ra\la s_{25}\w s_{24}\ra}{\la s_{35}\w s_{25}\ra\la s_{23}\w s_{24}\ra}
\,,\\
x_4&=e^{X_4}=\frac{\langle h_{0\fb}\xi_{21}\wedge h_{0\fb}\xi_{23}\rangle\langle h_{0\fb}\xi_{25}\wedge  h_{0\fb}G_{23}\xi_{35}\rangle}{\langle  h_{0\fb}\xi_{21}\wedge h_{0\fb}\xi_{25}\rangle\langle h_{0\fb}\xi_{23}\wedge  h_{0\fb} G_{23}\xi_{35}\rangle}
=\f{\la s_{21} \w s_{23}\ra\la s_{25}\w s_{35}\ra}{\la s_{21}\w s_{25}\ra\la s_{23}\w s_{35}\ra}\,.
\end{aligned}\ee
Therefore, denoting the spinor for cusp $i$ of $\cS_a$ based at $\fb_2$ as $\xi_i^{(2)}$, we have
\be
e^{\chi_{14}^{(2)}}=x_1x_4
=\f{\la s_{21} \w s_{23}\ra\la s_{25}\w s_{24}\ra}{\la s_{21}\w s_{25}\ra\la s_{23}\w s_{24}\ra}
=\f{[ \xi^{(2)}_4 | O^{-1}_3\xi^{(2)}_{1}\ra[ \xi^{(2)}_{3}| \xi^{(2)}_{2}\ra}{[ \xi^{(2)}_{4}| O^{-1}_3\xi^{(2)}_{3}\ra[ \xi^{(2)}_{1}| \xi^{(2)}_{2}\ra}\,,
\ee
which matches the result in \eqref{eq:FG_spheres}. The appearance of $O_3^{-1}$ comes from the path parallel transported $s_{ab}$ from $\fb_0$ to $\fb$ along different paths in defining $x_1$ and $x_4$. In particular, the former path has to be chosen to be within $\Oct(1)$ while the latter is within $\Oct(4)$. Other FG coordinates $\chi^{(a)}_{bc}$ can be calculated with the 10 spinors \eqref{eq:all_spinors} similarly. 
In this way, $\{\ell^2_{ab}=e^{2L_{ab}}\}_{a<b}, \{e^{M_a},e^{P_a}\}_a$ reconstructed in Section \ref{subsec:FG_on_Sa} can all be reproduced.  

It remains to check the consistency of $\tau_{ab}$ reconstructed in \eqref{eq:tau_chi} and $\{\fp_I\}_{I=1,\cdots,10}$ in \eqref{eq:q_p_from_phi_pi}. 
It relies on the fact that the result in \eqref{eq:tau_chi} is deduced from the ``snake rule" on the cusp \cite{Dimofte:2011gm}, which gives their expressions in terms of $\{\Phi_I,\Pi_I\}_I$ (see \eqref{eq:T}). We again examine it with an example: $T_{25}=\f12(Y'_1-X'_1)$, which is calculated in $\Oct(1)$ ({\it r.f.} fig.\ref{fig:oct1}). As $\fb$ is in $\Oct(1)$, one can directly use the set of spinors \eqref{eq:all_spinors} to calculate
\be\begin{split}
\tau_{25}^2=\f{y'_1}{x'_1}&=\f{\la G_{24}\xi_{45}\w G_{23}\xi_{35}\ra \la \xi_{25}\w \xi_{24}\ra}{\la G_{24}\xi_{45}\w\xi_{25}\ra \la G_{23}\xi_{35}\w\xi_{24}\ra}\cdot
\f{\la \xi_{24}\w G_{23}\xi_{35}\ra \la \xi_{23}\w \xi_{25}\ra}{\la \xi_{24} \w \xi_{23}\ra \la G_{23}\xi_{35}\w \xi_{25}\ra}\\
&=-\f{\la \xi_{23}\w \xi_{25}\ra\la \xi_{24}\w \xi_{25}\ra }{\la \xi_{23} \w \xi_{24}\ra}\cdot
\f{\la G_{24}\xi_{45}\w G_{23}\xi_{35}\ra  }{\la G_{24}\xi_{45}\w\xi_{25}\ra\la G_{23}\xi_{35}\w \xi_{25}\ra}\\
&=-\f{\la \xi_{23}\w \xi_{25}\ra\la \xi_{24}\w \xi_{25}\ra }{\la \xi_{23} \w \xi_{24}\ra}\cdot
\f{\la \gamma_{45}^{-1}G_{24}G_{45}\xi_{54}\w \gamma_{35}^{-1}G_{23}G_{35}\xi_{53}\ra  }{\la \gamma_{45}G_{24}G_{45}\xi_{54}\w \gamma_{25}G_{25}\xi_{52}\ra\la \gamma_{35}G_{23}G_{35}\xi_{53}\w \gamma_{25}\Gamma_{25}\xi_{52}\ra}\\
&=-\gamma_{25}^2\f{\la \xi_{23}\w \xi_{25}\ra\la \xi_{24}\w \xi_{25}\ra }{\la \xi_{23} \w \xi_{24}\ra}\cdot
\f{\la G_{25}\xi_{54}\w G_{25}\xi_{53}\ra  }{\la G_{25}\xi_{54}\w G_{25}\xi_{52}\ra\la G_{25}\xi_{53}\w G_{25}\xi_{52}\ra}\\
&=-\gamma_{25}^2\f{\la \xi_{23}\w \xi_{25}\ra\la \xi_{24}\w \xi_{25}\ra }{\la \xi_{23} \w \xi_{24}\ra}\cdot
\f{\la \xi_{54}\w \xi_{53}\ra  }{\la \xi_{54}\w \xi_{52}\ra\la \xi_{53}\w \xi_{52}\ra}\,,
\end{split}\ee
where we have used \eqref{eq:xi_G_relation} in the third line for the second cross-ratio and the $\SL(2,\bC)$ invariance property of the inner product $\la \cdot \w\cdot\ra=\la g \cdot \w g\,\cdot\ra \,(\forall\, g\in\SL(2,\bC))$ to obtain the last line. The first cross-ratio in the last line is defined at $\fb$ on $\cS_2$ while the second cross-ratio is defined at $\fb_5$ on $\cS_5$. The above result matches exactly that in \eqref{eq:tau_chi} when $(ab)=(25)$ as $\gamma_{25}=-\f12\nu\sgn(V_4)\Theta_{25}+i\theta_{25}$. Other FN twists can be calculated similarly.

Therefore, all the coordinate elements $\{\fq_I,\fp_I\}_I$ can be calculated either using spinors based on different 4-holed spheres or spinors based on a common based point on $\SG$, and they give the same results. 

\medskip

Finally, let us comment on the consistency of the geometrical reconstruction of the coordinates on $\cM_{\Flat}(\SG,\SL(2,\bC))$ described in this section and the 4-simplex geometry described using holonomies language described in Section \ref{sec:flat_connection_2_geometry}. In particular, we demonstrate here that the closure conditions \eqref{eq:closure_5spheres} and the parallel transport relations \eqref{eq:constraints_on_Oab} of holonomies have been used implicitly in the geometrical reconstruction using spinors and spins. 

First of all, the closure conditions \eqref{eq:closure_5spheres} are the gauge-fixed versions of the defining relations \eqref{eq:closure_3manifold} for the fundamental groups of both a 4-simplex and $\SG$ due to the isomorphism \eqref{eq:iso_simplex_manifold}. By defining FG coordinates on 4-holed spheres using spinors and spins, the holonomies $\{O_{ab}\}$ constructed with these coordinates of $\cM_\Flat(\Sigma_{0,4},\SU(2))\subset\cM_\Flat(\SG,\SL(2,\bC))$ (using the snake rules \cite{Dimofte:2011gm}) satisfy the closure conditions by definition. 
Moreover, we have applied the parallel transport relations \eqref{eq:xi_G_relation} between spinors in the geometry reconstruction for coordinates on 4-holed spheres in Section \ref{subsec:FG_on_Sa}, which was also implicitly used in Section \ref{subsec:coord_on_SG} by using only 10 spinors \eqref{eq:all_spinors}. (It eliminates $\{\xi_{ba}\}_{a<b}$ using \eqref{eq:xi_G_relation}.) 

For a general 4-complex dual to a colorable spinfoam graph, the above geometry reconstruction can be directly applied as the phase space coordinates describing the critical points are simply the linear combinations of the coordinates for one single 4-simplex as described above. We refer to \cite{Han:2025abc2} for more details on this point.

\section{Conclusion and discussion}
\label{sec:conclusion}

In this paper, we have demonstrated how to use geometrical data to reconstruct the critical points of the 4D Lorentzian spinfoam model with a nonzero $\Lambda$ constructed in \cite{Han:2025abc2}. 
The spinfoam model is built from the $\SL(2,\bC)$ Chern-Simons theory with a complex coupling constant on the graph-complement $\SG$, which is closely related to a 4-simplex. 
The real part of the coupling constant $k$ is taken to be 8 times a positive integer.
By mapping the geometry of a constantly curved 4-simplex to moduli space of $\SL(2,\bC)$ flat connection on $\SG$, which is the solution space to the Chern-Simons theory, we rebuild the Chern-Simons phase space coordinates using the geometry information of the 4-simplex. These coordinates contribute to the critical points of the vertex amplitude. The geometrical reconstruction is performed on 4-complices whose spinfoam graphs are colorable that are used in the colored tensor models, as defined in \cite{Han:2025abc2}. 

It gives an algorithm to calculate the critical points of the spinfoam amplitude with $\Lambda\neq 0$. 
	As minimum input, consider 5 distinct and independent points $v_1,\cdots,v_5$ on a $S^3$ (when $\Lambda>0$) or on a $\bH^3$ (when $\Lambda<0$), which identically define a non-degenerate constantly curved 4-simplex. Given the coordinates of all these points, one can calculate the areas and normals of all the triangles of the 4-simplex, yeilding an over-complete set of geometrical data. These data allow us to construct the FG and FN coordinates using the geometrical reconstruction as described in Section \ref{sec:geo_reconstruct}, which form the critical points of the spinfoam amplitude, leading to the zeroth-order spinfoam amplitude. 
	
	Starting from the zeroth-order amplitude, one can perform perturbations on the critical points, systematically generating next-to-leading order and higher-order quantum gravity corrections in the spinfoam framework. 
	 This algorithm is applicable to any 4-complex covered by the study of colored tensor model, and can be implemented numerically, as has been done in EPRL model \cite{Han:2020fil}. 
	 A key advantage of the spinfoam model with $\Lambda\neq 0$ over the case with $\Lambda=0$ is its finiteness property. 
	 Since the model remains finite for any 4-complex, numerical computations can be performed with higher accuracy and efficiency, without requiring artificial truncations. This feature makes it particularly promising for obtaining reliable quantum corrections in spinfoam models. We leave the numerical realization of this program for future investigation. 

It is interesting to investigate how this algorithm applies to physical systems such as cosmology and black holes. Attempts have been made using the EPRL model \cite{Han:2024ydv,Han:2024rqb}. 
In particular, the spinfoam model with a nonzero $\Lambda$ is naturally suited for studying quantum corrections to classical cosmology, which aligns with observational data from the current universe. To apply our approach in a cosmological setting, the spinfoam model should be coupled to a scalar field and adapted to a triangulation that respects the homogeneity and isotropy of spacetime. Using the geometrical construction of critical points and perturbative expansion, we expect to derive an effective Friedmann equation with a bare $\Lambda$ in the semiclassical regime. This would provide a consistency check for the covariant approach to loop quantum gravity by comparing the result with predictions from loop quantum cosmology with a nonzero $\Lambda$ \cite{Pawlowski:2011zf}.

\begin{acknowledgements}
The author thanks Muxin Han for insightful discussions. This work receives support from the Blaumann Foundation and the Jumpstart Postdoctoral Program at Florida Atlantic University. 
\end{acknowledgements}

\appendix
\renewcommand\thesection{\Alph{section}}

\section{Symplectic coordinates on $\partial(\SG)$}
\label{app:symplec_tranf}

In this appendix, we give detailed expressions on the FN coordinates in terms of the coordinates $(\vec{\cQ},\vec{\cP})$ on the ideal octahedra before the symplectic transformations. 

The FG coordinates $\{\chi_{bc}^{(a)}\}_{b,c}$ and coordinates $\{\{L_{ab}\}_{b},M_a,P_a\}$ are related by linear combination with integer prefactors as follows. 
\be
\ba{lll}
\chi^{(1)}_{23}=-L_{12}+L_{14}+P_1\,,\quad
&\chi^{(1)}_{24}=-L_{14}+L_{15}+M_1\,,\quad
&\chi^{(1)}_{25}=L_{13}+L_{14}-M_1-P_1+3 i \pi \,,\\[0.15cm]
\chi^{(1)}_{34}=L_{12}+L_{15}-M_1-P_1+3 i \pi \,,\quad
&\chi^{(1)}_{35}=L_{12}-L_{13}+M_1\,,\quad
&\chi^{(1)}_{45}=L_{13}-L_{15}+P_1\,,\\[0.35cm]
\chi^{(2)}_{13}=L_{12}+L_{25}+M_2+P_2\,,\quad
&\chi^{(2)}_{14}=L_{23}+L_{25}-M_2+3 i \pi\,,\quad
&\chi^{(2)}_{15}=L_{24}-L_{25}-P_2\,,\\[0.15cm]
\chi^{(2)}_{34}=-L_{12}-L_{23}-P_2\,,\quad
&\chi^{(2)}_{35}=-L_{12}+L_{24}-M_2+3 i \pi\,,\quad
&\chi^{(2)}_{45}=L_{23}-L_{24}+M_2+P_2\,,\\[0.35cm]
\chi^{(3)}_{12}=L_{34}+L_{35}-P_3+3 i \pi\,,\quad
&\chi^{(3)}_{14}=-L_{23}-L_{34}-M_3+P_3\,,\quad
&\chi^{(3)}_{15}=L_{13}+L_{34}+M_3\,,\\[0.15cm]
\chi^{(3)}_{24}=L_{23}+L_{35}+M_3\,,\quad
&\chi^{(3)}_{25}=-L_{13}-L_{35}-M_3+P_3\,,\quad
&\chi^{(3)}_{45}=-L_{13}-L_{23}-P_3+3 i \pi\,,\\[0.35cm]
\chi^{(4)}_{12}=L_{14}-L_{34}+M_4\,,\quad
&\chi^{(4)}_{13}=L_{34}+L_{45}+P_4\,,\quad
&\chi^{(4)}_{15}=-L_{24}-L_{34}-M_4-P_4+3 i \pi \,,\\[0.15cm]
\chi^{(4)}_{23}=-L_{14}+L_{45}-M_4-P_4+3 i \pi\,,\quad
&\chi^{(4)}_{25}=-L_{14}+L_{24}+P_4\,,\quad
&\chi^{(4)}_{35}=-L_{24}-L_{45}+M_4\,,\\[0.35cm]
\chi^{(5)}_{12}=L_{25}-L_{35}+M_5-P_5\,,\quad
&\chi^{(5)}_{13}=-L_{25}-L_{45}-M_5+3 i \pi\,,\quad
&\chi^{(5)}_{14}=L_{15}-L_{25}+P_5\,,\\[0.15cm]
\chi^{(5)}_{23}=L_{35}-L_{45}+P_5\,,\quad
&\chi^{(5)}_{24}=-L_{15}-L_{35}-M_5+3 i \pi\,,\quad
&\chi^{(5)}_{34}=-L_{15}+L_{45}+M_5-P_5\,.
\ea
\label{eq:chi_to_LMP}
\ee
The FN lengths $2L_{ab}$ and FN twists $T_{ab}$ are calculated using the snake rule on cusps \cite{Dimofte:2011gm} \footnote{A good reference for computing $T_{ab}$ to get the results \eqref{eq:T} is Appendix H of \cite{Han:2023hbe}, noting that the role of $\fz'_i$ and $\fz''_i$ ($\fz_i=x_i,y_i,z_i,w_i$) are exchanged therein compared to this paper.}.
Explicitly,
\be
\begin{aligned}
&  2L_{12}=\chi^{(1)}_{34}+\chi^{(1)}_{35}+\chi^{(1)}_{45}-3i\pi=-P_{Y_3}-P_{Y_4}-P_{Y_5}-Y_3-Y_4-Y_5-Z_3-Z_4-Z_5+3 i \pi \,,\\
&  2L_{13}=\chi^{(1)}_{24}+\chi^{(1)}_{25}+\chi^{(1)}_{45}-3i\pi=-P_{Y_2}-P_{Y_4}+P_{Z_4}-P_{Z_5}-Y_2-Y_4-Z_2+Z_4+i \pi\,, \\
&  2L_{14}=\chi^{(1)}_{23}+\chi^{(1)}_{25}+\chi^{(1)}_{35}-3i\pi=-P_{Y_2}-P_{Y_5}+P_{Z_2}-P_{Z_3}+P_{Z_5}-Y_2-Y_5+Z_2+Z_5\,, \\
&  2L_{15}=\chi^{(1)}_{23}+\chi^{(1)}_{24}+\chi^{(1)}_{34}-3i\pi=-P_{Y_3}-P_{Z_2}+P_{Z_3}-P_{Z_4}-Y_3+Z_3\,, \\
&  2L_{23}=\chi^{(2)}_{14}+\chi^{(2)}_{15}+\chi^{(2)}_{45}-3i\pi=P_{X_1}-P_{X_4}+P_{X_5}+P_{Y_4}+2 X_1+2 X_5+Y_1+Y_5+Z_1+Z_5-4 i \pi\,, \\
&  2L_{24}=\chi^{(2)}_{13}+\chi^{(2)}_{15}+\chi^{(2)}_{35}-3i\pi=P_{X_3}-P_{X_5}+P_{Y_1}+P_{Y_5}+2 X_3+Y_1+Y_3+Z_1+Z_3-3 i \pi\,, \\
&  2L_{25}=\chi^{(2)}_{13}+\chi^{(2)}_{14}+\chi^{(2)}_{34}-3i\pi=-P_{X_1}-P_{X_3}+P_{X_4}+P_{Y_1}+P_{Y_3}+2 X_4+Y_4+Z_4-2 i \pi \,, \\
&  2L_{34}=\chi^{(3)}_{12}+\chi^{(3)}_{15}+\chi^{(3)}_{25}-3i\pi=-P_{X_2}-P_{X_5}+P_{Y_2}+P_{Z_1}+P_{Z_5}+Y_5+Z_5-i \pi \,, \\
&  2L_{35}=\chi^{(3)}_{12}+\chi^{(3)}_{14}+\chi^{(3)}_{24}-3i\pi=-P_{X_1}+P_{X_2}+P_{X_4}+P_{Z_1}-P_{Z_4}+2 X_2+Y_1+Y_2-Y_4+Z_1+Z_2-Z_4-2 i \pi\,, \\
&  2L_{45}=\chi^{(4)}_{12}+\chi^{(4)}_{13}+\chi^{(4)}_{23}-3i\pi=P_{X_2}-P_{X_3}-P_{Y_1}+P_{Z_1}-P_{Z_2}+P_{Z_3}-Y_1-Y_2+Y_3+Z_1-Z_2+Z_3\,.
\end{aligned}
\label{eq:L}
\ee
\be
\begin{aligned}
&  T_{12}=\frac{1}{2} \left(-X'_3+Y_3-Z''_3\right)=\frac{1}{2} \left(P_{X_3}-P_{Z_3}+X_3+Y_3-i \pi \right)\,\,
&&  T_{13}=\frac{1}{2} \left(W_5-Z'_5\right)=\frac{1}{2} \left(P_{Z_5}+X_5+Y_5+2 Z_5-2 i \pi \right)\,, \\
&  T_{14}=\frac{1}{2} \left(Z''_2-Y''_2\right)=\frac{1}{2} \left(P_{Z_2}-P_{Y_2}\right) \,,\,
&&  T_{15}=\frac{1}{2} \left(W_4-Z'_4\right)=\frac{1}{2} \left(P_{Z_4}+X_4+Y_4+2 Z_4-2 i \pi \right)\,, \\
&  T_{23}=\frac{1}{2} \left(Y'_4-X'_4\right)=\frac{1}{2} \left(P_{X_4}-P_{Y_4}+X_4-Y_4\right)\,, \,
&&  T_{24}=\frac{1}{2} \left(Y'_5-X'_5\right)=\frac{1}{2} \left(P_{X_5}-P_{Y_5}+X_5-Y_5\right)\,, \\
&  T_{25}=\frac{1}{2} \left(Y'_1-X'_1\right)=\frac{1}{2} \left(P_{X_1}-P_{Y_1}+X_1-Y_1\right) \,, \,
&&  T_{34}=\frac{1}{2} \left(Z'_1-W_1\right)=\frac{1}{2} \left(-P_{Z_1}-X_1-Y_1-2 Z_1+2 i \pi \right)\,, \\
&  T_{35}=\frac{1}{2} \left(X''_2-W''_2\right)=\f12P_{X_2}\,, \,
&&  T_{45}=\frac{1}{2} \left(W_3+Y''_3-Z_3\right)=\frac{1}{2} \left(P_{Y_3}+X_3+Y_3-i \pi \right)\,.
\end{aligned}
\label{eq:T}
\ee

\section{Calculate the Fock-Goncharov coordinates with spinors and holonomies}
\label{app:FG_calculation}

In this appendix, we give an example of how to calculate the FG coordinates using spinors on a 4-holed sphere $\cS_a$. In particular, we derive $x_{12}$ in \eqref{eq:FG_spheres} and verify the first equality in \eqref{eq:chi_to_2L}. 

\begin{figure}
\centering
\includegraphics{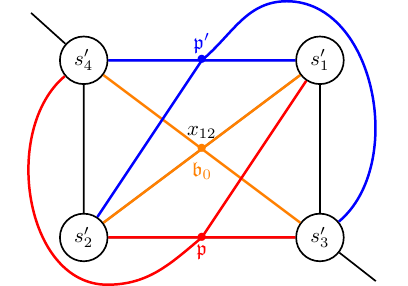}
\caption{$s'_i$ $(i=1,\cdots,4)$ is the framing flag located at cusp $i$ as the eigenvector of holonomy $O_i$ and $O'_i$ surronding cusp $i$ with eigenvalue $e^{\pm 2\pi i\frac{j_i}{k}}$. $O_i$ starts at $\fp$ on the edge connecting cusps 2 and 3 while $O'_i$ starts at $\fp'$ on the edge connecting cusps 1 and 4. The parallel transported framing flag $s_i$ of $s'_i$ to $\fb_0$ on the edge connecting cusps 1 and 2 are used to calculate FG coordinate $x_{12}$ through \eqref{eq:FG_from_flag}. }
\label{fig:frame_flag_FG_comb}
\end{figure}

Refering to fig.\ref{fig:frame_flag_FG_comb}, the FG coordinate $x_{12}$ is calculated with framing flags $s_i$ ($i=1,\cdots,4$), which are framing flags $s'_i$ on cusps parallel transported to a common point, chosen to be $\fb_0$ located on the edge connecting cusps 1 and 2, inside the parallelogram. Denote the holonomy from cusp $i$ to $\fb_0$ as $h_{0i}$ along the {\it orange} line, the holonomy from cusp $i$ to $\fp$ (\resp $\fp'$) as $h_{\fp i}$ along the {\it red} line (\resp $h_{\fp'i}$ along the {\it blue} line), and $h_{ij}^{-1}\equiv h_{ji}$. 
The role of framing flags can be played by the spinors. Denote the spinor parallel transported from cusp $i$ to $\fp$ (\resp $\fp'$) as $\xi_i$ (\resp $\zeta_i$). Then 
\be
\xi_i=h_{\fp i}s'_i=h_{\fp i}h_{i0}s_i\,,\quad
\zeta_i=h_{\fp'i}s'_i=h_{\fp'i}h_{i0}s_i\,.
\ee
\eqref{eq:FG_from_flag} for $x_{12}$ can then be formulated in terms of $\xi_i$ or $\zeta_i$ as follows
\begin{align}
x_{12}=
\frac{\left\langle s_4 \wedge s_2\right\rangle\left\langle s_1 \wedge s_3\right\rangle}{\left\langle s_4 \wedge s_1\right\rangle\left\langle s_2 \wedge s_3\right\rangle}
=\frac{\left[ h_{04}h_{4\fp}\xi_4 | h_{02}h_{2\fp}\xi_2\right\rangle\left[ h_{01}h_{1\fp}\xi_1 | h_{03}h_{3\fp}\xi_3\right\rangle}{\left[ h_{04}h_{4\fp}\xi_4 | h_{01}h_{1\fp}\xi_1\right\rangle\left[ h_{02}h_{2\fp}\xi_2 | h_{03}h_{3\fp}\xi_3\right\rangle}
&=\frac{\left[ \xi_4 | h_{\fp4}h_{40}h_{02}h_{2\fp}\xi_2\right\rangle\left[ \xi_1 | h_{\fp1}h_{10}h_{03}h_{3\fp}\xi_3\right\rangle}{\left[ \xi_4 | h_{\fp4}h_{40}h_{01}h_{1\fp}\xi_1\right\rangle\left[ \xi_2 | h_{\fp2}h_{20}h_{03}h_{3\fp}\xi_3\right\rangle}\nn\\
&\equiv\f{[\xi_4|O_2\xi_2\ra[\xi_1|\xi_3\ra}{[\xi_4|O_2\xi_1\ra[\xi_2|\xi_3\ra}\,,\\
x_{12}=
\frac{\left\langle s_4 \wedge s_2\right\rangle\left\langle s_1 \wedge s_3\right\rangle}{\left\langle s_4 \wedge s_1\right\rangle\left\langle s_2 \wedge s_3\right\rangle}
=\frac{\left[ h_{04}h_{4\fp'}\zeta_4 | h_{02}h_{2\fp'}\zeta_2\right\rangle\left[ h_{01}h_{1\fp'}\zeta_1 | h_{03}h_{3\fp'}\zeta_3\right\rangle}{\left[ h_{04}h_{4\fp'}\zeta_4 | h_{01}h_{1\fp'}\zeta_1\right\rangle\left[ h_{02}h_{2\fp'}\zeta_2 | h_{03}h_{3\fp'}\zeta_3\right\rangle}
&=\frac{\left[ \zeta_4 | h_{\fp'4}h_{40}h_{02}h_{2\fp'}\zeta_2\right\rangle\left[ \zeta_1 | h_{\fp'1}h_{10}h_{03}h_{3\fp'}\zeta_3\right\rangle}{\left[ \zeta_4 | h_{\fp'4}h_{40}h_{01}h_{1\fp'}\zeta_1\right\rangle\left[ \zeta_2 | h_{\fp'2}h_{20}h_{03}h_{3\fp'}\zeta_3\right\rangle}\nn\\
&\equiv\f{[\zeta_4|\zeta_2\ra[\zeta_1|O'^{-1}_1\zeta_3\ra}{[\zeta_4|\zeta_1\ra[\zeta_2|O'^{-1}\zeta_3\ra}\,.
\end{align}

Other formulas in \eqref{eq:FG_spheres} can be obtained in the same way. Using these results, let us also verify the first equality of \eqref{eq:chi_to_2L}:
\be\begin{split}
x_{12}x_{23}x_{24}&=\f{[\xi_4|O_2\xi_2\ra[\xi_1|\xi_3\ra}{[\xi_4|O_2\xi_1\ra[\xi_2|\xi_3\ra}
\f{[\xi_1|\xi_2\ra[\xi_3|\xi_4\ra}{[\xi_1|\xi_3\ra[\xi_2|\xi_4\ra}
\f{[\xi_1|O_2^{-1}\xi_4\ra[\xi_2|\xi_3\ra}{[\xi_1|O_2^{-1}\xi_2\ra[\xi_4|\xi_3\ra}\\
&=-\frac{[\xi_{4}|O_{2}\xi_{2}\rangle}{[\xi_{2}|O_{2}\xi_{1}\rangle}\frac{[\xi_{1}|\xi_{2}\rangle}{[\xi_{2}|\xi_{4}\rangle}\\
&=-\frac{\left([\xi_{4}|\xi_{2}\rangle,[\xi_{4}|\xi_{2}]\right)\left(\begin{array}{cc}
 e^{-2\pi i\frac{j_{2}}{k}} & 0\\
0 &  e^{2\pi i\frac{j_{2}}{k}}
\end{array}\right)\left(\begin{array}{c}
\langle\xi_{2}|\xi_{2}\rangle\\{}
[\xi_{2}|\xi_{2}\rangle
\end{array}\right)}{\left([\xi_{2}|\xi_{2}\rangle,[\xi_{2}|\xi_{2}]\right)\left(\begin{array}{cc}
 e^{-2\pi i\frac{j_{2}}{k}} & 0\\
0 & e^{2\pi i\frac{j_{2}}{k}}
\end{array}\right)\left(\begin{array}{c}
\langle\xi_{2}|\xi_{1}\rangle\\{}
[\xi_{2}|\xi_{1}\rangle
\end{array}\right)}\frac{[\xi_{1}|\xi_{2}\rangle}{[\xi_{2}|\xi_{4}\rangle}\\
&=-\frac{\left([\xi_{4}|\xi_{2}\rangle,[\xi_{4}|\xi_{2}]\right)\left(\begin{array}{cc}
 e^{-2\pi i\frac{j_{2}}{k}} & 0\\
0 &  e^{2\pi i\frac{j_{2}}{k}}
\end{array}\right)\left(\begin{array}{c}
1\\0
\end{array}\right)}{\left(0,1\right)\left(\begin{array}{cc}
 e^{2\pi i\frac{j_{2}}{k}} & 0\\
0 & e^{-2\pi i\frac{j_{2}}{k}}
\end{array}\right)\left(\begin{array}{c}
\langle\xi_{2}|\xi_{1}\rangle\\{}
[\xi_{2}|\xi_{1}\rangle
\end{array}\right)}\frac{[\xi_{1}|\xi_{2}\rangle}{[\xi_{2}|\xi_{4}\rangle}\\
&=-\frac{e^{-2\pi i\frac{j_{2}}{k}}[\xi_{4}|\xi_{2}\rangle}{e^{2\pi i\frac{j_{2}}{k}}[\xi_{2}|\xi_{1}\rangle}\frac{[\xi_{1}|\xi_{2}\rangle}{[\xi_{2}|\xi_{4}\rangle}=-e^{-4\pi i\frac{j_{2}}{k}}\,,
\end{split}\ee
where we have used the properties of $\SU(2)$ spinors: $[\xi_i|\xi_j\ra=-[\xi_j|\xi_i\ra$, $\la \xi_i|\xi_i\ra=[\xi_i|\xi_i]=1$, $[\xi_i|\xi_i\ra=0$ and the factorization \eqref{eq:O_factorize} of $O_2$. Other equalities in \eqref{eq:chi_to_2L} can be verified in the similar way.

\section{Geometry encoded Fock-Goncharov coordinates on ideal tetrahedra and ideal octahedra}
\label{app:FG_on_Oct}

In this appendix, we collect the explicit expressions of the FG coordinates on ideal tetrahedra of $\TSG$ in terms of the 10 spinors \eqref{eq:all_spinors} encoding the normal vector to triangles in a 4-simplex and the dihedral angles $\Theta_{21},\Theta_{23},\Theta_{24}$ between tetrahedra in a 4-simplex. The FG coordinates on $\Oct(1)$ are 
\be
\begin{aligned}
&x_1=\frac{\langle G_{23}\xi_{35}\wedge\xi_{23}\rangle\langle\xi_{25}\wedge\xi_{24}\rangle}{\langle G_{23}\xi_{35}\wedge\xi_{25}\rangle\langle\xi_{23}\wedge\xi_{24}\rangle}\,,\quad 
x'_1=\frac{\langle\xi_{23}\wedge\xi_{24}\rangle\langle\xi_{25}\wedge G_{23}\xi_{35}\rangle}{\langle\xi_{23}\wedge\xi_{25}\rangle\langle\xi_{24}\wedge G_{23}\xi_{35}\rangle}\,,\quad 
x''_1=\frac{\langle\xi_{23}\wedge\xi_{25}\rangle\langle G_{23}\xi_{35}\wedge\xi_{24}\rangle}{\langle\xi_{23}\wedge G_{23}\xi_{35}\rangle\langle\xi_{25}\wedge\xi_{24}\rangle}\,,\\
&y_1=\frac{\langle G_{23}\xi_{35}\wedge\xi_{25}\rangle\langle G_{24}\xi_{45}\wedge\xi_{24}\rangle}{\langle G_{23}\xi_{35}\wedge G_{24}\xi_{45}\rangle\langle\xi_{25}\wedge\xi_{24}\rangle}\,,\quad 
y'_1=\frac{\langle\xi_{25}\wedge\xi_{24}\rangle\langle G_{24}\xi_{45}\wedge G_{23}\xi_{35}\rangle}{\langle\xi_{25}\wedge G_{24}\xi_{45}\rangle\langle\xi_{24}\wedge G_{23}\xi_{35}\rangle}\,,\quad 
y''_1=\frac{\langle\xi_{25}\wedge G_{24}\xi_{45}\rangle\langle G_{23}\xi_{35}\wedge\xi_{24}\rangle}{\langle\xi_{25}\wedge G_{23}\xi_{35}\rangle\langle G_{24}\xi_{45}\wedge\xi_{24}\rangle}\,,\\
&z_1=\frac{\langle\xi_{24}\wedge G_{23}\xi_{34}\rangle\langle G_{24}\xi_{45}\wedge G_{23}\xi_{35}\rangle}{\langle\xi_{24}\wedge G_{24}\xi_{45}\rangle\langle G_{23}\xi_{34}\wedge G_{23}\xi_{35}\rangle}\,,\,\, 
z'_1=\frac{\langle G_{23}\xi_{34}\wedge G_{23}\xi_{35}\rangle\langle G_{24}\xi_{45}\wedge\xi_{24}\rangle}{\langle G_{23}\xi_{34}\wedge G_{24}\xi_{45}\rangle\langle G_{23}\xi_{35}\wedge\xi_{24}\rangle}\,,\,\, 
z''_1=\frac{\langle G_{23}\xi_{34}\wedge G_{24}\xi_{45}\rangle\langle\xi_{24}\wedge G_{23}\xi_{35}\rangle}{\langle G_{23}\xi_{34}\wedge\xi_{24}\rangle\langle G_{24}\xi_{45}\wedge G_{23}\xi_{35}\rangle}\,,\\
&w_1=\frac{\langle G_{23}\xi_{35}\wedge G_{23}\xi_{34}\rangle\langle\xi_{23}\wedge\xi_{24}\rangle}{\langle G_{23}\xi_{35}\wedge\xi_{23}\rangle\langle G_{23}\xi_{34}\wedge\xi_{24}\rangle}\,,\quad 
w'_1=\frac{\langle G_{23}\xi_{34}\wedge\xi_{24}\rangle\langle\xi_{23}\wedge G_{23}\xi_{35}\rangle}{\langle G_{23}\xi_{34}\wedge\xi_{23}\rangle\langle\xi_{24}\wedge G_{23}\xi_{35}\rangle}\,,\quad
w''_1=\frac{\langle G_{23}\xi_{34}\wedge\xi_{23}\rangle\langle G_{23}\xi_{35}\wedge\xi_{24}\rangle}{\langle G_{23}\xi_{34}\wedge G_{23}\xi_{35}\rangle\langle\xi_{23}\wedge\xi_{24}\rangle}\,.
\end{aligned}
\label{eq:FG_Oct1}
\ee
They automatically satisfy the constraint defining $\cL_{\Oct(1)}$:
\be
x_1x_1'x_1''=y_1y_1'y_1''=z_1z_1'z_1''=w_1w_1'w_1''=-1\,,\quad
x_1''+x_1^{-1}=y_1''+y_1^{-1}=z_1''+z^{-1}=w_1''+w_1^{-1}=1\,,\quad
x_1y_1z_1w_1=1\,.
\ee
One then define the conjugate momenta $p_{x_1},p_{y_1},p_{y_1}$ to $x_1,y_1,z_1$ respectively by
\be
p_{x_1}:=x''_1-w''_1\,,\quad
p_{y_1}:=y''_1-w''_1\,,\quad
p_{z_1}:=z''_1-w''_1\,.
\ee
The corresponding FG coordinates on $\Oct(i)$ with $i=2,\cdots,5$ are defined as in \eqref{eq:FG_Oct1} with a change of spinors as follows. 
\be
\ba{lllllll}
\Oct(2): &\xi_{23}\rightarrow G_{23}\xi_{34}\,,&
\xi_{24}\to	G_{21}\xi_{13}\,,&
\xi_{25}\to	G_{23}\xi_{35}\,,&
G_{23}\xi_{34}\to	G_{21}\xi_{14}\,,&
G_{23}\xi_{35}\to	G_{24}\xi_{45}\,,&
G_{24}\xi_{45}\to	G_{21}\xi_{15}\,,\\[0.15cm]
\Oct(3): &\xi_{23}	\to\xi_{24}\,,&
\xi_{24}\to	\xi_{21}\,,&
\xi_{25}\to	\xi_{25}\,,&
G_{23}\xi_{34}\to	G_{21}\xi_{14}\,,&
G_{23}\xi_{35}\to	G_{24}\xi_{45}\,,&
G_{24}\xi_{45}\to	G_{21}\xi_{15}\,,\\[0.15cm]
\Oct(4): &\xi_{23}	\to\xi_{23}\,,&
\xi_{24}\to	\xi_{21}\,,&
\xi_{25}\to	\xi_{25}\,,&
G_{23}\xi_{34}\to	G_{21}\xi_{13}\,,&
G_{23}\xi_{35}\to	G_{23}\xi_{35}\,,&
G_{24}\xi_{45}\to	G_{21}\xi_{15}\,,\\[0.15cm]
\Oct(5): &\xi_{23}	\to\xi_{23}\,,&
\xi_{24}\to	\xi_{21}\,,&
\xi_{25}\to	\xi_{24}\,,&
G_{23}\xi_{34}\to	G_{21}\xi_{13}\,,&
G_{23}\xi_{35}\to	G_{23}\xi_{34}\,,&
G_{24}\xi_{45}\to	G_{21}\xi_{14}\,.
\ea\ee

\bibliographystyle{bib-style} 
\bibliography{SFC.bib}

\end{document}